\documentclass[reprint,aps,twocolumn,prb,superscriptaddress,nofootinbib]{revtex4-2}

\usepackage{indentfirst,csquotes}
\usepackage{amsmath}
\usepackage{bm}
\usepackage[english]{babel}
\usepackage{graphicx,psfrag}
\usepackage{amsfonts}
\usepackage{tikz}
\usetikzlibrary{shapes,arrows}
\usepackage{lipsum}
\usepackage{physics}
\usepackage[colorlinks=true]{hyperref}
\hypersetup{
allcolors=blue}
\usepackage[utf8]{inputenc}
\usepackage{amssymb}
\usepackage{natbib}
\usepackage{mathtools}
\usepackage{appendix}
\usepackage{comment}
\usepackage{marvosym}
\usepackage{subcaption}
\usepackage[flushleft]{threeparttable}

\newcommand{\psibar}{\bar{\psi}}

\newcommand{\Aomega}{\mathcal{A}_{\omega}}
\newcommand{\Ak}{\mathcal{A}_{\mathbf{k}}}
\newcommand{\BAomega}{\mathcal{B}_{\mathbf{A},\Omega}}
\newcommand{\BAq}{\mathcal{B}_{\mathbf{A},\mathbf{q}}}
\newcommand{\Bsq}{\mathcal{B}_{\phi,\mathbf{q}}}
\newcommand{\kUV}{k_{\text{UV}}}
\newcommand{\GammaL}{\Gamma^\Lambda}
\newcommand{\qbar}{\bar{q}}

\newcommand{\etaAomega}{\eta_{\mathbf{A},\Omega}}
\newcommand{\etaAq}{\eta_{\mathbf{A},\mathbf{q}}}

\newcommand{\etabar}{\bar{\eta}}

\newcommand{\defeq}{\vcentcolon=}

\setcitestyle{numbers,square}

\begin{document}

\title{Non-Fermi liquid induced by U(1) gauge field interactions: a functional renormalization group analysis}
\author{Thomas P. Sheerin}
\affiliation{SUPA, School of Physics and Astronomy, University of St Andrews,
North Haugh, St Andrews, Fife KY16 9SS, United Kingdom}

\author{Chris A. Hooley}
\affiliation{Centre for Fluid and Complex Systems, Coventry University, Coventry CV1 2TT, United Kingdom}

\begin{abstract}
We study the non-Fermi-liquid state formed by an isotropic, degenerate Fermi gas in two spatial dimensions interacting with a U(1) gauge field.  Our calculation uses the functional renormalization group (fRG) with a soft frequency cutoff for the fermions. The fRG scheme we employ takes account of the gauge symmetry, which imposes relations (modified Ward-Takahashi identities) between the couplings which constrain the RG flow. The critical exponents and couplings we find for the resulting non-Fermi liquid are mostly insensitive to whether or not we enforce the gauge symmetry constraints, which signifies either that the constraints are superfluous or that the frequency-cutoff scheme is particularly robust. The exception is the gauge-boson mass term, which is RG-relevant about the fixed point without the constraints, but is irrelevant when they are enforced. The latter is physically accurate, as a gauge symmetry cannot be spontaneously broken. In addition, we find $z=2$ and $\Sigma(\omega,k_F)\sim\omega^{1/2}$ for the boson dynamical exponent and scaling of the fermion self-energy, respectively. These results differ considerably from those of past works on the model, though we argue for their plausibility.
\end{abstract}

\maketitle

\section{Introduction}

A full theoretical understanding of non-Fermi liquids (NFLs), metallic states that have no fermionic quasiparticle excitations, remains a key challenge to modern theoretical physics~\cite{SSBK2011,SL2018}, one which concerns various subfields. In the condensed matter community, NFLs are believed to occur in a number of material families (ruthenates~\cite{SGRPAS2001,GCOKSC2008}, iron-based superconductors~\cite{TSACYM2014,EAQS2011}, heavy fermion compounds~\cite{JCPGHW2003,FGIWSJ2001,HLTPGP1994} and cuprates~\cite{SS2010a,SMAFRF1990,RDNDDL2009}) and could act as ``parent states'' from which to launch studies of exotic ordered phases, like high-$T_c$ superconductivity~\cite{AS1999}. In the high-energy community, on the other hand, they could provide insight into dense quark-gluon plasmas, wherein interactions between quarks and transverse gauge fields can lead to NFL behavior~\cite{TS2007,TSKS2004}. Even aside from these specific fields of interest, the canonical picture of a NFL, that of fermions with a Fermi surface destabilized by a gapless boson, is a paradigmatic problem in field theory and the theory of the renormalization group (RG), one which is still quite poorly understood, particularly in spatial dimension $d=2$~\cite{SL2018,SL2009,STWM2011,AAAC2004,MMSS2010b}.

This gap in knowledge arises due to various difficulties uniquely inherent to modelling NFLs, which cause the usual tools used to study similar problems to fail. For example, the Hertz-Millis~\cite{JH1976,AM1993} approach of entirely ``integrating out'' (following a Hubbard-Stratonovich transformation) the fermions in favour of the bosonic order parameter has been shown to break down in $d=2$~\cite{AAAC2004,STWM2011} due to singular bosonic self-interactions arising from integrating over gapless fermion modes; another recourse, the limit of large fermion-flavour number, was shown to be pathological in various important universality classes~\cite{SL2009,MMSS2010b}. Other techniques, such as quantum Monte Carlo~\cite{YLWJAK2022,YSSLSK2016}, holography~\cite{SL2009holo,SHALSS2018,HLJMDV2011} and more sophisticated RG approaches~\cite{AFSKJK2013,SMPS2016,MTCH2018,CNFW1994}, have been employed with far more success in recent years. However, since so little is known about the form of the low-energy theories of most NFLs, most approaches must make some uncontrolled approximations in order to make headway. This results in a loss of predictive power and surety, and might cause key features to be missed. An example particularly pertinent to this paper is that of the functional RG, for which truncation and parametrization of the effective action are not controlled by a small parameter and may be ill-motivated. 

An approach that has garnered much interest of late is that of using exact identities that survive renormalization in order to constrain modelling procedures, or check their validity. For example, 't Hooft anomaly matching has been shown to place strong constraints on the low-energy physics, even in settings without quasiparticles~\cite{DERTTS2021,CWAHXY2021,XW2021,DE2024}. Another source of constraints lies in symmetries~\cite{YLWJAK2022,MMSS2010a,FSLBPK2005,ACJBDE2014,AC2005}, since these manifest in field theory as exact relations between correlation functions, which is particularly tantalizing for assessing the accuracy of RG schemes. Gauge symmetries are particularly restrictive, and so in order to gain insight into the effect of symmetry constraints on RG procedures for NFLs one could study a system with a continuous gauge symmetry, the simplest of which is U(1).

The model of fermions at finite density interacting with a U(1) gauge field in two dimensions has been proposed as a low-energy effective field theory for many physically relevant systems, such as the strange metal phase of the cuprate high-$T_c$ superconductors\ \cite{JP1994,PLNNXW2006,GBPA1988,ANTM1988,PL1989}, spin liquids\ \cite{SLPL2005,XW1989,OM2005}, and electrons in a half-filled Landau level\ \cite{BHPLNR1993,YKPLXW1995}.
Additionally, it is well known that the Fermi liquid is destabilized by interactions with the gauge field. Decades ago it was found~\cite{THRNPP1973,MR1989,MR1991,PL1989} that, though the Coulomb field is screened, the magnetic vector potential $\mathbf{A}$ is unscreened by the particle-hole continuum and causes the fermionic quasiparticle weight to vanish. Since then the model has been treated by a variety of techniques~\cite{BBHM1993,OMMF2007,MMSS2010a,SL2009,IM2020,JP1994,SCRNOS1995,THWM2015,CNFW1994,DKPS1993,YKAFXW1994,BALIAM1994,BHPLNR1993}. Most of these~\cite{BBHM1993,OMMF2007,MMSS2010a,SL2009,IM2020,JP1994,SCRNOS1995,THWM2015,DKPS1993,YKAFXW1994,BALIAM1994,BHPLNR1993} capture Landau damping by first performing one-loop perturbation theory for the inverse gauge-boson propagator (or equivalently, using the RPA), producing a $\propto |\Omega|/|\mathbf{q}|$ term (generated by the particle-hole bubble), and then proceeding with other methods like RG. A dynamical exponent $z_b=3$ results. However, this is an unsatisfactory, ``Hertz-Millis'' style approach in that all fermionic degrees of freedom, including the gapless ones at the Fermi surface, contribute with the same weight to a coupling at the start of the flow, which contravenes the RG philosophy of sequentially integrating out modes, starting at the highest energies. At the level of the action, the problem is manifested in the $\propto |\Omega|/|\mathbf{q}|$ term, which is non-analytic and so should not develop at an intermediate stage in the flow. An additional unsatisfactory feature of most of the above approaches is that they are perturbative in coupling constants, which is not well suited to $d=2$, where NFLs often occur at $\mathcal{O}(1)$ couplings. 

The functional RG with a soft frequency cutoff function for the fermion field is a consistent approach which allows Landau damping to develop gradually during the flow --- this will be delineated in Subsec.~\ref{subsec:props} (the form we use was developed in Ref.~\cite{SMPS2016}). The functional RG also allows the design of schemes that are non-perturbative (i.e.\ not relying on small couplings, though not necessarily exact).

In this work, we study the model of a U(1) gauge field interacting with an isotropic, degenerate Fermi gas at zero temperature, in two spatial dimensions, treated using the functional RG. We assume the presence of either time-reversal ($\mathcal{T}$) or parity ($\mathcal{P}$) symmetry --- were these both absent a Chern-Simons term would be allowed for the gauge field~\cite{JP1994}, which would presumably give a different universality class. We use a soft frequency regulator to allow Landau damping to develop in a consistent way, and correctly take account of the model's gauge symmetry through (modified) Ward-Takahashi identities. In order to assess the extent to which these identities change the low-energy physics, we perform the analysis twice, once with the identities enforced and once without. In addition, we do not treat the effect of instantons in our analysis. Whether or not they prevent a deconfined phase for compact, two-dimensional U(1) gauge theory in the presence of a Fermi surface is unclear~\cite{YZKLYW2012,IHBSSS2003,SL2008,KK2005,GAIITM2006}, but we assume they do not (see Subsec.\ \ref{subsec:gen_form}). Our analysis thus may or may not be relevant for compact U(1), but is certainly applicable to non-compact U(1).

The remainder of the paper is organized as follows. In Sec.~\ref{sec:not_conv} we briefly summarize our notation, then in Sec.~\ref{sec:frg_symms} we recapitulate the essentials of the functional RG and how symmetries are incorporated in it. In Sec.~\ref{sec:Gamma}, we parametrize the effective average action of the low-energy theory, and in Sec.~\ref{sec:mWTIs} calculate the modified Ward-Takahashi identities it obeys. In Sec.~\ref{sec:method} we explain how we enforce the identities alongside the RG flow equations, and in Sec.~\ref{sec:results} we present our results for the model, along with some discussion. We conclude in Sec.~\ref{sec:conclusion}.

\section{Notation and Conventions}
\label{sec:not_conv}

Though it is expected that the Coulomb potential of the gauge field is strongly screened and does not affect the universal features, we initially include it in the analysis as some of its features prove instructive. We therefore treat the gauge field $A_\mu$ in full, as in standard quantum electrodynamics. Relativistic notation will occasionally appear as a result: bosonic three-momenta are $q^\mu = (i\Omega,\mathbf{q})$ and the metric is $g^{\mu\nu}=\text{diag}(+1,-1,-1)$. The bosonic and fermionic integration measures are, respectively, 

\begin{equation}
\int_q \defeq \int_{-\infty}^\infty \frac{d\Omega}{2\pi}\int_{-\infty}^{\infty} \frac{dq_x}{2\pi}\int_{-\infty}^{\infty}\frac{dq_y}{2\pi}~,
\end{equation}
and
\begin{equation}
\int_k \defeq k_F\int_{-\infty}^\infty \frac{d\omega}{2\pi}\int_{-\kUV}^{\kUV} \frac{dl}{2\pi}\int_0^{2\pi}\frac{d\theta}{2\pi}~.
\end{equation}

In the latter, $l=|\mathbf{k}|-k_F$ is the perpendicular distance from the Fermi surface, $\kUV$ is a UV momentum cutoff and the measure $|\mathbf{k}| dl$ has been replaced by $k_F dl$ as the difference is irrelevant under the RG~\cite{RS1994}. The dimensionless ratio $N=k_F/\kUV$ will appear in the low-energy theory (this non-universality is discussed in Sec.~\ref{sec:results}). Finally, we will denote the fully dressed fermion propagator by $G$, and the gauge-field propagator by $D$.

\section{The Functional RG and Gauge Symmetries}
\label{sec:frg_symms}

Before considering the specifics of our model, let us briefly summarize the functional renormalization group (fRG), and how the choice of scheme interplays with symmetries of the action. 

The fRG is an implementation of the RG framework that is derived by considering the flow of a theory's generating functionals, which yields a flow equation for the effective average action $\Gamma^\Lambda$ which is formally exact (before various neccessary approximations are implemented). The great power of this approach is that it facilitates approximations that are non-perturbative in coupling constants and that can handle non-analytic dependences on frequencies and momenta in a natural way. Specifically, the Wetterich flow equation reads~\cite{WMMSCH2012,CW1993}

\begin{equation}
\label{eqn:Wetterich}
\partial_\Lambda \Gamma^\Lambda = \frac{1}{2}\text{Str}\left[\partial_\Lambda\mathbf{R}^\Lambda\left(\mathbf{\Gamma}^{(2)\Lambda}+\mathbf{R}^\Lambda\right)^{-1}\right]~.
\end{equation}
Here $\mathbf{R}^\Lambda=\text{diag}(R^\Lambda_f,-(R^\Lambda_f)^T,R^\Lambda_b)$, where $R^\Lambda_{f/b}$ are the fermion/boson regulator functions and are added to the action to regularize divergences and induce flow. Each should be thought of as an infinite-dimensional matrix, with entries labelled by spacetime and internal degrees of freedom. $\mathbf{\Gamma}^{(2)\Lambda}$ is the matrix of second derivative of $\Gamma^\Lambda$ with respect to the fermion and boson fields, and the supertrace Str traces over all indices of its argument, with a minus sign in the fermionic sector. Writing $\Gamma^\Lambda$ as a ``Maclaurin series'' in the fields (with coefficients equal to the 1PI correlation functions), substituting into the above and comparing like orders yields an exact hierarchy of one-loop flow equations, one for each of the 1PI vertex functions. Thus far the regulator functions need only satisfy a few properties (aside from ensuring the IR and UV divergences are cut off, $\mathbf{R}^{\Lambda\to0}$ must also vanish), but later we shall make choices that are particularly appropriate for describing Landau damping.

Now, consider a theory with fermion fields $\psi_{\uparrow,\downarrow}$ which possesses a U(1) gauge symmetry with associated gauge field $A_\mu$. To facilitate conceptual separation of the Coulomb field and vector potential, we work in Coulomb gauge {$\nabla\cdot\mathbf{A}=\mathbf{0}$}, which corresponds to addition to the action of a gauge-fixing term

\begin{align}
S_\text{gf}=\frac{1}{2 \xi}\int_q(\mathbf{q}\cdot\mathbf{A}(q))(\mathbf{q}\cdot\mathbf{A}(-q))~,
\end{align}
where we set the gauge fixing parameter $\xi$ to 0 after finding the propagator. The full, unregularized theory satisfies the master Ward-Takahashi identity (WTI)
\begin{align}
&\frac{1}{\xi}\mathbf{q}^2\mathbf{q}\cdot\mathbf{A}(q)-q_{\mu}\frac{\delta \Gamma}{\delta A_{\mu}(-q)}\nonumber\\&-e\int_k\left[\psibar(k-q)\frac{\delta\Gamma}{\delta\psibar(k)}+\frac{\delta \Gamma}{\delta\psi(k)}\psi(k+q)\right]=0.
\end{align}

Expanding $\Gamma$ in a series as before one arrives at an infinite family of WTIs. Specifically, if $(\Gamma^{(2m,n)})^{\mu_1...\mu_n}$ denotes the 1PI vertex function with $n$ gauge-field lines and $m$ incoming fermions, then there is a WTI relating $q_{\mu} (\Gamma^{(2m,n+1)})^{\mu\mu_1...\mu_n}$ to a linear expression in $(\Gamma^{(2m,n)})^{\mu_1...\mu_n}$ (at various arguments). For example:
\begin{align}
\label{eqn:(2,1)WTI}&q_\mu \left(\Gamma^{(2,1)}\right)^\mu(k-q,k,-q)=\nonumber\\&-e\left(\Gamma^{(2,0)}(k)-\Gamma^{(2,0)}(k-q)\right)~,
\end{align}
where $e$ is the unrenormalized charge. This is an exact relation between renormalized correlation functions, and as such is an attractive prospect for designing robust approximation schemes for $\Gamma$. 

\begin{figure}
\centering
\includegraphics[width=\linewidth]{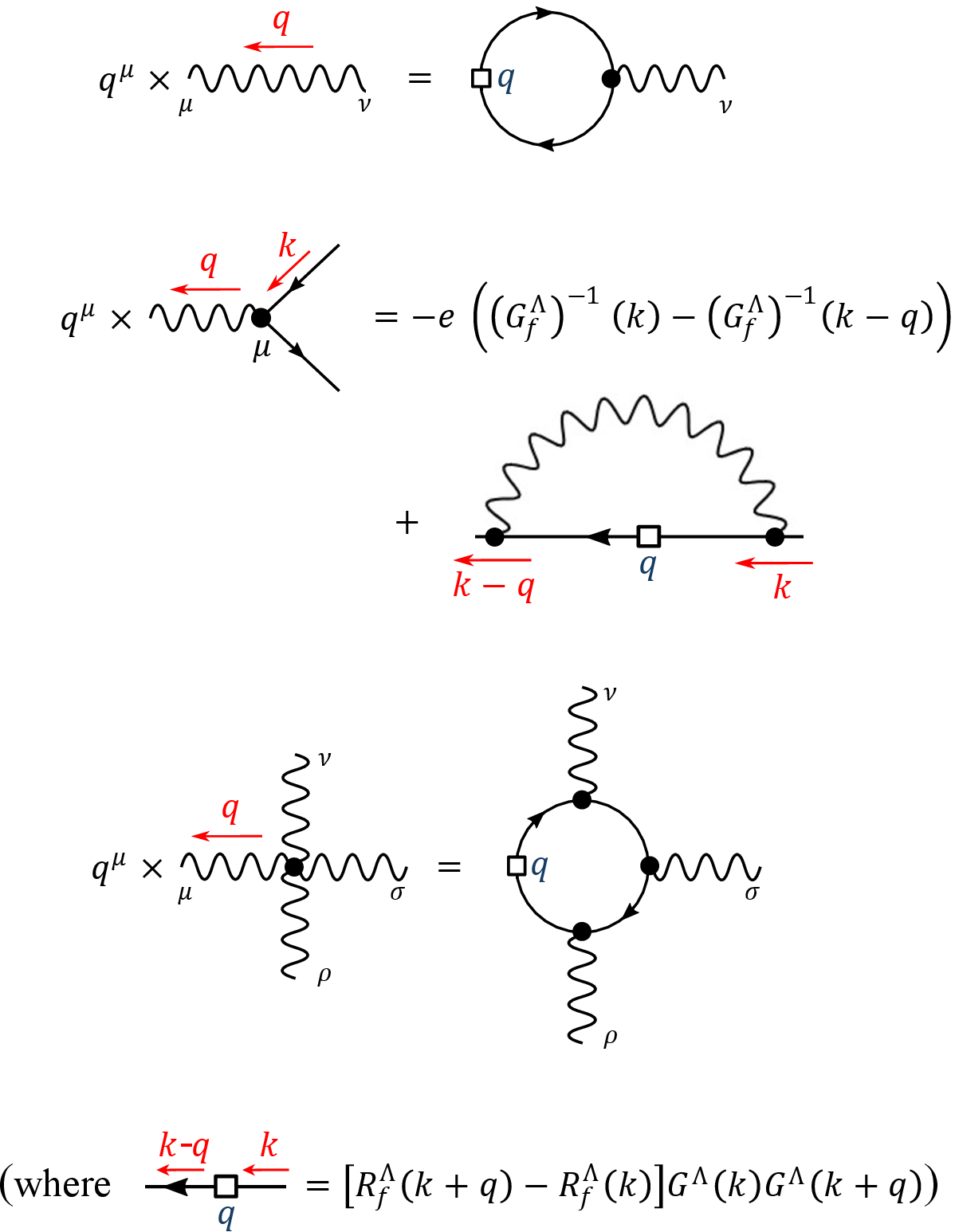}
\caption{Diagrammatic modified Ward-Takahashi identities for the model. The boxed lines represent the propagator $[R^\Lambda_{f}(k+q)-R^\Lambda_{f}(k)]G^\Lambda(k)G^\Lambda(k+q)$. In the third identity the diagram with the boxed fermionic propagator represents the sum of three diagrams, in each of which the box is placed on a different one of the fermionic propagator lines.}
\label{fig:mWTIs}
\end{figure}

Addition of regulators generically breaks the gauge symmetry, leading to so-called modified Ward-Takahashi identities (mWTIs). The master identity is

\begin{align}
\label{eqn:mastermWTI}
&\frac{1}{\xi}\mathbf{q}^2\mathbf{q}\cdot\mathbf{A}(q)-q_{\mu}\frac{\delta \Gamma^\Lambda}{\delta A_{\mu}(-q)}\nonumber\\&-e\int_k\left[\psibar(k-q)\frac{\delta\Gamma^\Lambda}{\delta\psibar(k)}+\frac{\delta \Gamma^\Lambda}{\delta\psi(k)}\psi(k+q)\right]=\nonumber\\&e\int_k\left(R_{f,k+q}^{\Lambda}-R_{f,k}^{\Lambda}\right)\sum\limits_{\sigma}\frac{\delta^2 \mathcal{G}^{\Lambda}}{\delta \eta_{\sigma,k}\delta \etabar_{\sigma,k+q}}\Biggr\vert_{\eta^*,\etabar^*,J^*}~
\end{align}
(the momenta have been placed in subscript in the final line for readability). Here, $\eta,\etabar$ and $J$ are sources for $\psi,\psibar$ and $A$ respectively, and $\mathcal{G}^\Lambda$ is the generating functional for connected correlation functions (the Legendre transform of the average action). $\eta^*,\etabar^*,J^*$ are then the Legendre-transform fields, satisfying $\psibar = \delta \mathcal{G}^\Lambda/\delta \eta \rvert_{\eta^*,\etabar^*,J^*}$, etc. It is apparent that the above is the usual WTI term (which usually = 0) set equal to a one-loop modification term. The gauge field regulator makes no appearance, and as such can be chosen quite freely. This is a feature peculiar to an abelian gauge symmetry: in the SU$(N)$ mWTIs, there is a further term on the right above, proportional to the structure factors~\cite{HG2012}. One can again expand $\GammaL$ and derive a hierarchy of identities --- the modification to the identity for $q_\mu (\Gamma^{(2m,n+1)\Lambda})^{\mu\mu_1...\mu_n}$ is found by summing all one-loop diagrams with $(2m,n)$ external lines and at least one internal fermion line, with single insertions of the propagator $[R^\Lambda_f(k+q)-R^\Lambda_f(k)]G^\Lambda(k)G^\Lambda(k+q)$ on each of the internal fermion lines in turn ($G^\Lambda$ being the fermion propagator). It is apparent, from this, that at the end of the flow $R^\Lambda_f\to 0$ and the usual WTIs are recovered, i.e.\ gauge symmetry is restored. The identities pertinent to this work are shown in Fig.~\ref{fig:mWTIs}, and as an example~(\ref{eqn:(2,1)WTI}) is modified to
\begin{widetext}
\begin{eqnarray}
\label{eqn:Yukawa_mWTI}
q_\mu \left( \Gamma^{(2,1)\Lambda}\right)^\mu(k-q,k,-q) & = & -e\left(\Gamma^{(2,0)\Lambda}(k)-\Gamma^{(2,0)\Lambda}(k-q)\right)+ e\int_{k'}\left[R^\Lambda_f(k'+q)-R^\Lambda_f(k')\right]G^\Lambda(k')G^\Lambda(k'+q)\times\nonumber \\
& & \qquad \times D^{\Lambda \mu\nu}(k'-k+q)\Gamma_{\mu}^{(2,1)\Lambda}(k'+q,k,k'-k+q) \Gamma_{\nu}^{(2,1)\Lambda}(k-q,k',k-q-k')~. 
\end{eqnarray}
The remaining identities are given in Appendix~\ref{append:deriv}. Their one-loop structure makes the mWTIs somewhat more cumbersome to deal with than the WTIs, but in this model they prove tractable. 

\section{Effective Average Action Ansatz}
\label{sec:Gamma}
\subsection{General form}
\label{subsec:gen_form}
Let us now parametrize an ansatz for the low-energy effective average action $\Gamma^\Lambda$. Henceforth we shall drop all superscripts $\Lambda$ on flowing quantities for notational clarity.  Our ansatz takes the form
\begin{align}
\label{eqn:Ansatz}
\Gamma[\psi,\psibar,A]=&-\int_k \sum_\sigma\psibar_\sigma(k)\left[G^{-1}(k)-R_f(k)\right]\psi_\sigma(k)+\frac{1}{2}\int_q A_\mu(q)\left[(D^{-1})^{\mu\nu}(q)-R_A^{\mu\nu}(q)\right]A_\nu(-q)\nonumber\\&-\int_{k,q}\sum_\sigma\Gamma^{(2,1)\mu}(k+q,k,q)\psibar_\sigma(k+q)\psi_\sigma(k)A_\mu(q)\nonumber\\&+\frac{1}{4!}\int_{q_1,q_2,q_3} \Gamma^{(0,4)\mu\nu\rho\sigma}(q_1,q_2,q_3,-q_1-q_2-q_3)A_\mu(q_1) A_\nu(q_2)A_\rho(q_3)A_\sigma(-q_1-q_2-q_3)~.
\end{align}
\end{widetext}
 All higher-order vertices are irrelevant under the scaling we introduce below, save $\Gamma^{(0,5)}$ and $\Gamma^{(0,6)}$:\ we neglect these purely for reasons of analytic tractability.  A boson three-point function $\Gamma^{(0,3)}$ has not been included, as it would generically make the transition to the ordered state first-order.  We are not aware of any microscopic argument showing that such a term must vanish, since non-relativistic electrodynamic models usually do not have the charge-conjugation symmetry enjoyed by full quantum electrodynamics.  Hence, if such a symmetry obtains in the low-energy theory, it must be emergent. Here 
we \emph{assume} the transition is second order, and leave the investigation of the correctness of this assumption to future work. We have also manually pinned the four-Fermi interaction to zero; this is common in investigations of NFLs~\cite{MTCH2018,SMPS2016} since it prevents pairing instabilities and allows access to the critical point. Finally, we have also omitted a $\Gamma^{(2,2)}$ vertex (such a term is present in the UV action of nonrelativistic QED) --- we have tried including this and found it does not affect the low-energy physics.

A comment on the possible influence of instantons is in order. U(1) gauge theory can be ``compact'' or ``non-compact'', for which the gauge group is U(1) or ($\mathbb{R}$,+), respectively. The perturbative structures of the theories are identical, and all of the terms included in the effective action here and below are permitted in either. However, the topological structure of the gauge group can make a drastic difference to the ground state. It was shown by Polyakov~\cite{AP1977} that compact, pure QED (i.e.\ without coupling to matter) in two spatial dimensions is always confining due to the effect of instantons. In the three-dimensional Euclidean picture these instantons are to be thought of as monopoles charged under the magnetic U(1) symmetry, and their long-range interactions give a mass to the photon of that symmetry, which in turn causes confinement.

Whether or not this conclusion holds in the presence of charged fermions~\cite{IHBS2003,FNHK2008,MHTSMF2004,WASHJK2011}, and particularly when those fermions have a Fermi surface~\cite{YZKLYW2012,IHBSSS2003,SL2008,KK2005,GAIITM2006}, is actively debated. The effect of instantons may be included through the addition to the effective action of terms involving ``instanton creation operators''~\cite{VBBHCH2021,SL2008}, so that the presence or absence of confinement is determined by whether or not these terms survive under the RG. We shall assume that there exists a region of the phase diagram in which they are RG-irrelevant --- there is some evidence to support this~\cite{SL2008,KK2005}, though the existence of a deconfined phase is definitely contentious~\cite{IHBSSS2003,YZKLYW2012}. Our analysis thus may or may not be applicable to compact U(1), but is certainly valid for non-compact U(1).

\subsection{Yukawa couplings}
Now, as is standard~\cite{AFSKJK2013,jp1999}, we perform a derivative expansion of the coupling functions and retain terms based on their scaling dimensions about the Gaussian fixed point. Choosing momenta and frequencies to scale as $[\omega]=[\Omega]=[l]=[|\mathbf{q}|]=1$, the field scaling dimensions must be 
\begin{equation}
    [\psi_\sigma]=-\frac{3}{2}~,\hspace{4mm}[A_\mu]=-\frac{5}{2}~.
\end{equation}
Then, $\Gamma^{(2,1)}$ is irrelevant at large momentum transfers parallel to the Fermi surface, while for small ones it has scaling dimension $1/2$~\cite{AFSKJK2013}. Expanding in $l,q_x,q_y,\omega$ and $\Omega$, we find that only the constant term is non-irrelevant. Note that this term may still depend on the angle $\mathbf{k}$ makes to some reference direction --- indeed, it must, as $\Gamma^{(2,1)\mu}(\hat{\mathbf{k}})$ must transform appropriately under SO$(2)$. Specifically, $\Gamma^{(2,1)0}$ is a scalar (say $-\tilde{g}_\phi$) and $\Gamma^{(2,1) i}$ a vector. The latter implies
\begin{equation}
\label{eqn:Yukawa_form}
\Gamma^{(2,1)i}(\hat{\mathbf{k}})=-\tilde{g}_\mathbf{A} \left(R_{\theta_0}\hat{\mathbf{k}}\right)^i~,
\end{equation}
where $R_{\theta_0}$ is some fixed rotation matrix. Consistency with the mWTIs will imply $\theta_0=0$ (see Appendix~\ref{append:deriv}).

\subsection{Propagators}
\label{subsec:props}

The fermion propagator is
\begin{align}
G(k)=\chi(\omega,\Lambda)\frac{1}{i\Aomega\omega-\Ak l}~,
\end{align}
where we choose the ``soft'' cutoff function
\begin{equation}
\chi(\omega,\Lambda)=\frac{\omega^2}{\omega^2+\Lambda^2}~.
\end{equation}
This identifies $\Lambda$ as a frequency cutoff. This form, used in Refs.~\cite{SMPS2016,MTCH2018}, is well suited to describing Landau damping in a ``natural'' way. Small-momentum particle-hole excitations, which generate the Landau damping, are included early in the flow and not suppressed, as they would be in hard-momentum-cutoff schemes~\cite{SMPS2016,KGMS2012}. On the other hand, it avoids the unsatisfactory procedure of dressing the boson propagator by one-loop corrections (given by the particle-hole bubble) at the beginning of the RG flow, as is done in Hertz-Millis theory~\cite{JH1976,AM1993}. Such a procedure entails unregularized integration over all gapless fermion modes, giving rise to a non-analytic $|\Omega|/|\mathbf{q}|$ term in the effective average action which ought not to develop at intermediate scales in a consistent RG scheme; our effective average action remains local.

Non-trivial dynamical scaling should still be possible at the end of the flow: we capture it as follows. In the above propagator, $\Aomega$ and $\Ak$ are flowing RG variables (they correspond to quasiparticle weight $\mathcal{Z}_{\text{qp}}=1/\Aomega$ and Fermi velocity $v_F=\Ak/\Aomega$), and at criticality scale as $\Lambda^{-\eta_\omega}$ and $\Lambda^{-\eta_\mathbf{k}}$, respectively. Interpreting these anomalous scalings as telling us that the powers of momentum and frequency in the propagator must be adjusted, we have, on shell,
\begin{equation}
\Lambda^{-\eta_\omega} \Lambda \sim \Lambda^{-\eta_\mathbf{k}}\Lambda^{1/z_f} \quad
\Leftrightarrow \quad z_f=\frac{1}{1-\eta_\omega+\eta_\mathbf{k}}~,
\end{equation}
and analogously for the boson.

Next, the gauge field. In Coulomb gauge, we have
\begin{widetext}
\begin{equation}
D^{ij}(q)=-\frac{1}{\BAomega\Omega^2+\BAq|\mathbf{q}|^2+M_\mathbf{A}^2+R_\mathbf{A}}\left(g^{ij}+\frac{q^i q^j}{\mathbf{q}^2}\right), \quad
D^{00}(q)=-\frac{1}{\Bsq |\mathbf{q}|^2+M_\phi^2+R_\phi}, \quad D^{0i}=D^{i0}=0~.
\end{equation}
\end{widetext}
Here, $\BAomega,\BAq$ and $\Bsq$ renormalize the dependences on frequency and momenta (with flowing $\mathbf{A}$-field speed $c^2=\BAq/\BAomega$), $R_{\mathbf{A}/\phi}$ are regulators (see below) and $M_{\mathbf{A}/\phi}$ are mass terms, which are allowed by the mWTIs. The above propagators arise from the usual Maxwell kinetic term, the simplest gauge-invariant form (up to the mass terms) consistent with the $\mathcal{T}$ and $\mathcal{P}$ symmetries that we assume for the low-energy theory.  Relaxing both of these requirements would allow a Chern-Simons term~\cite{JP1994}; we do not consider that case here. 

The above corresponds to a self-energy
\begin{align}
\label{eqn:Pi_Ansatz}
\Pi^{\mu\nu}(q)&=\Pi_\mathbf{A}(q)\left(g^{\mu\nu}-\left(\delta_0^\mu\delta_0^\nu-\frac{\qbar^\mu\qbar^\nu}{\mathbf{q}^2}\right)\right)\nonumber\\&+\Pi_\phi(q)\left(\frac{q^\mu q^\nu}{\mathbf{q}^2}+\frac{\Omega^2+\mathbf{q}^2}{\mathbf{q}^2}\left(\delta_0^\mu\delta_0^\nu-\frac{\qbar^\mu\qbar^\nu}{\mathbf{q}^2}\right)\right)\nonumber\\&+M_\phi^2 \delta^\mu_0\delta^\nu_0+M_\mathbf{A}^2(g^{\mu\nu}-\delta^\mu_0\delta^\nu_0)~,
\end{align}
with $\qbar^\mu=(0,\mathbf{q})$. The first two lines give the most general transverse part (consistent with $\mathcal{T}$- and $\mathcal{P}$-invariance), while the last is only allowed by the regularization scheme's breaking gauge symmetry.

This latter term is not the most general addition possible, but it is a form commonly used in the high-energy literature~\cite{UEMHAW1995,HG2012} and we have strong reason to believe that it captures the critical physics.  This restriction brings with it, in principle, an ambiguity in how to match up the left- and right-hand sides of the flow equation for $\Pi^{\mu\nu}$; however, that is resolved once the scalar field is dropped, as we shall do.  We therefore choose, by convention, to extract the flow of the transverse part associated to $\phi$ by setting $\mu=\nu=0$ on the right-hand side, the flow of the transverse part associated to $\mathbf{A}$ by applying $(g_{ij}+q_iq_j/\mathbf{q}^2)$, and the flows of the longitudinal terms by applying $q_\mu$.

We then parametrize 
\begin{align}
\label{eqn:Pi_param}
    \Pi_\mathbf{A}(q)&=(\BAomega-1)\Omega^2+(\BAq-1)|\mathbf{q}|^2~,\nonumber\\
    \Pi_\phi(q)&=(\Bsq-1)|\mathbf{q}|^2~.
\end{align}
We could also include constant terms in these parametrizations, i.e.\ take $\Pi_{\mathbf{A}/\phi}(0)\neq 0$. These are mass terms that are allowed by gauge symmetry, as they multiply transverse projectors in~(\ref{eqn:Pi_Ansatz}). We neglect them because they would correspond to non-analytic operators in the effective action, which ought to be local at all intermediate scales. One could argue that the above ansatz~(\ref{eqn:Pi_Ansatz}) already contains very similar operators, like $\Omega^2~ \qbar^\mu \qbar^\nu/\mathbf{q}^2$ --- however, these are at least well behaved at $q=0$. Another possible objection is that the Coulomb field in metals is known to develop a gauge-invariant mass term through screening. This is addressed in Appendix~\ref{append:nonpert}.

For the $\mathbf{A}$-field regulator we make the simple choice $R_\mathbf{A}=\BAomega \Lambda^2$; we leave $R_\phi$ unspecified as we shall soon drop the Coulomb field.

\subsection{Boson four-point function}

We shall only keep $\Gamma^{(0,4)ijkl}(q_1,...,q_4)$ and neglect temporal indices, as we expect the $\mathbf{A}$ field to drive criticality, and will soon drop the $\phi$ field anyway. The four-point interaction has $[\Gamma^{(0,4)}]=1$, so we can keep the constant and linear-in-$q$ terms. However, the latter could only depend on frequency, so we omit them for simplicity. The most general form transverse to all the external momenta would be $\propto \varepsilon^{ii'}q_{1,i'}...\varepsilon^{ll'}q_{4,l'}$, $\varepsilon$ being the Levi-Civita symbol; however, to give this the proper scaling dimension, we would have to divide by some power of $|\mathbf{q}_1|...|\mathbf{q}_4|$, again giving the kind of non-analytic operator that shouldn't develop during the flow. We thus neglect it.

One finds that the mWTIs and flow equations also generate a term of the form $g^{ij}g^{lk}+g^{ik}g^{jl}+g^{il}g^{jk}$. This is non-transverse, and so must vanish at the end of the flow, but how it behaves at intermediate scales is still instructive. So, we parametrize as
\begin{align}
\label{eqn:lambda}
\Gamma^{(0,4)ijkl}&(q_1,...,q_4)=\nonumber\\&\tilde{\lambda} \left(g^{ij}g^{lk}+g^{ik}g^{jl}+g^{il}g^{jk}\right)~.
\end{align}

\section{Modified Ward-Takahashi identities}
\label{sec:mWTIs}

There are five identities; the first two are for the masses, and read
\begin{equation}
\label{eqn:mWTI_mass}
M_\phi^2=\frac{e\tilde{g}_\phi}{\pi \Ak}k_F\hspace{3mm},\hspace{3mm}M_\mathbf{A}^2=\frac{e\tilde{g}_\mathbf{A}N}{4\pi \Ak}\Lambda~.
\end{equation}
Recall that $N=k_F/\kUV$. The factor of $\Lambda$ in the identity for $M_\mathbf{A}^2$ ensures that it goes to zero at the end of the flow (see Subsec.~\ref{subsec:results_constraints} for discussion).  $M_\phi^2$ does not --- although it should, since gauge symmetry ought to be restored at the end of the flow, where the regulators vanish. We suspect that this is an artefact of the frequency-cutoff scheme, as the identity for $M_\phi^2$ arises from a frequency derivative of a diagram. It does not affect our analysis because we will soon neglect the Coulomb field; see Appendix~\ref{append:nonpert} for further discussion of this question. The next two identities, for the Yukawa couplings, are of the form
\begin{align}
\label{eqn:brief_Yukawa_mWTI}
&\tilde{g}_\phi=e\Aomega +\text{corrns.}~,\nonumber\\
&\tilde{g}_\mathbf{A}=e\Ak+\text{corrns.}~,
\end{align}
where ``corrns.'' denotes the one-loop corrections, which are given in Sec.~\ref{sec:method}. We find that the corrections to the identity for $\tilde{g}_\phi$ can sometimes be pathological, even becoming negative enough to force $\tilde{g}_\phi$ negative. This is another artefact of the frequency-cutoff scheme, but causes no difficulties since we drop the $\phi$ field. If we neglect the corrections to the above identities, we get the analogue of the classic QED WTI.

The mWTI for $\tilde{\lambda}$ is
\begin{equation}
    \tilde{\lambda} = \frac{5e \tilde{g}_v^3 N\Lambda}{64\pi\Ak^3 \kUV^2}~.
\end{equation}

\section{Method}
\label{sec:method}

Henceforth we shall drop the couplings associated to the $\phi$ field. We can confirm that inclusion of $\tilde{g}_s,\Bsq$ and $M_\phi^2$ does not change the properties of the critical point, at least within the limits of  our parametrization of $\Pi^{\mu\nu}$ and our choice of how to extract the associated flow equations --- the Coulomb field simply gets screened to a short-range interaction, as expected. There remain the couplings $\Aomega,\Ak,\BAomega,\BAq,M_\mathbf{A}^2,\tilde{g}_v,\tilde{\lambda}$. Let us transform to a set of dimensionless couplings $\Aomega,Y,\zeta,\kappa,\delta,g,\lambda$, with
\begin{align}
\label{eqn:rescaled_couplings}
Y=\frac{\Ak \kUV}{\Aomega \Lambda}~,~&\zeta=\frac{\Aomega \BAq^{1/2}}{\Ak \BAomega^{1/2}}~,\nonumber\\\kappa=\sqrt{\frac{N e^2}{8\pi \BAomega \Lambda}}~&,~\delta = \frac{M_\mathbf{A}^2}{\BAomega\Lambda^2}~,\nonumber\\g^2=\frac{N\tilde{g}_\mathbf{A}^2}{8\pi \Ak^2 \BAomega \Lambda}~,~&\lambda = \frac{\tilde{\lambda}}{8\pi \BAomega \BAq \Lambda}~.
\end{align}
Here, $Y=v_F\kUV/\Lambda$ may be assumed to diverge, as $v_F$ is not likely to go to zero as quickly as $\Lambda$. $\zeta=c/v_F$, the ratio of the fields' speeds, will tend to a constant $\ll 1$, consistent with overdamping of the gauge field. $\kappa$ is introduced simply to absorb the apparent dependence of the low energy theory on the highly non-universal initial coupling $e$ (this appearance is enforced by the mWTI). 

In terms of the above variables, the mWTIs take the much simpler forms 
\begin{align}
\label{eqn:mWTIs}
\delta = 2 g \kappa\hspace{2mm}&,\hspace{2mm}g=\kappa\left(1+\frac{1}{N}g^2 f(\delta,\zeta,Y)\right)~,\nonumber\\& \lambda=\frac{5\kappa g^3}{8 N \zeta^2 Y^2}~,
\end{align}
where $f$ is a complicated function defined by a double integral, given in Appendix~\ref{append:integrals}. The identities have been derived assuming $N\gg 1$ --- as we will argue in Sec.~\ref{sec:results}, this is the most physically accurate limit. Since $Y$ diverges and, as we will see, $\kappa,g$ and $\zeta$ go to non-zero constants at criticality, we note that $\lambda$ is quickly driven to zero, at least when the mWTIs are enforced. This is reassuring, since $\lambda$ multiplies the non-gauge-invariant coupling term~(\ref{eqn:lambda}), and so must vanish at the end of the flow when gauge symmetry is restored.

The beta functions for the couplings $Y,\zeta,\kappa,\delta,g$ and $\lambda$ are
\begin{widetext}
\begin{gather}
\beta_Y=Y(\eta_\omega-\eta_\mathbf{k}-1)~,\hspace{4mm}~\beta_\zeta
=\zeta\left(\frac{\etaAomega-\etaAq}{2}+\eta_\mathbf{k}-\eta_\omega\right)~,\hspace{4mm}~\beta_\kappa=\frac{\kappa}{2}\left(\etaAomega-1\right)~,\nonumber\\
\beta_g = g\left(\eta_\mathbf{k}-\frac{1}{2}+\frac{1}{2}\etaAomega-\eta_{\tilde{g}}\right)~,\nonumber\\\beta_\delta=3(2-\eta_\mathbf{k})g^2-\delta\left(2-\etaAomega\right)-\lambda(2-\etaAomega) \left[\frac{1}{\sqrt{\delta+1}}-\frac{1}{\sqrt{Y^2\zeta^2+\delta+1}}\right]~,\nonumber\\
\begin{align}\label{eqn:beta_functions}\beta_\lambda = \lambda \left(\etaAomega-\etaAq-1\right)+\frac{35 (2-\eta_\mathbf{k})}{16 Y^2}\frac{g^4}{\zeta^2 N}-\frac{9(\etaAomega-2)\lambda^2}{8} \left[\frac{1}{(\delta+1)^{3/2}}-\frac{1}{(Y^2\zeta^2+\delta+1)^{3/2}}\right]~,
\end{align}
\end{gather}
with
\begin{align}
\label{eqn:anom}
\eta_\omega=\frac{16\pi g^2}{N}\Bigr[-2i(\etaAomega-2)F_1(\delta,\zeta,Y)-&(2-\eta_\omega)F_2(\delta,\zeta,Y)-i(2-\eta_\mathbf{k})F_3(\delta,\zeta,Y)\Bigr]~,\nonumber\\
\eta_\mathbf{k}=-\frac{16\pi g^2}{N}\Bigr[(\etaAomega-2)F_4(\delta,\zeta,Y)-i(&2-\eta_\omega)F_5(\delta,\zeta,Y)+(2-\eta_\mathbf{k})F_6(\delta,\zeta,Y)\Bigr]~,\nonumber\\
\eta_{\tilde{g}}=\frac{8\pi g^2}{N}\Bigr[-(\etaAomega-2)G_1(\delta,\zeta,Y)+2i(&2-\eta_\omega)G_2(\delta,\zeta,Y)-2(2-\eta_\mathbf{k})G_3(\delta,\zeta,Y)\Bigr]~,\nonumber\\
\etaAomega = \frac{g^2}{2}&\left(1-\frac{\eta_\mathbf{k}}{2}\right)~,\nonumber\\
\etaAq = -\frac{g^2}{2\zeta^2 Y^2}&\left(1-\frac{\eta_\mathbf{k}}{2}\right)~,  
\end{align}
\end{widetext}
where the functions $F_i$ and $G_i$ are given in Appendix~\ref{append:integrals}. Again, the limit $N\gg 1$ has been assumed (see Appendix~\ref{append:deriv} for sample derivations). The anomalous dimension for $\Aomega$ is defined as
\begin{equation}
\eta_\omega = -\frac{\Lambda}{\Aomega}\partial_\Lambda \Aomega~,
\end{equation}
and the remaining four are for $\Ak,\tilde{g}_\mathbf{A},\BAomega$ and $\BAq$. The feedback of anomalous dimensions on the right hand side in~(\ref{eqn:anom}) complicates matters somewhat, but analytic solution is possible in the $N\to\infty$ limit.

\begin{figure}
\centering
\includegraphics[width=0.75\linewidth]{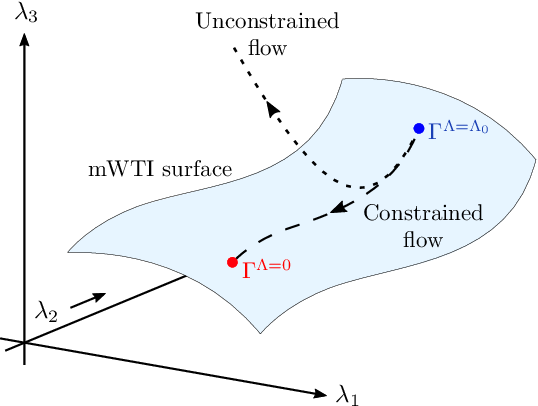}
\caption{Visualization of constraining in coupling space, for three couplings $\lambda_1,\lambda_2,\lambda_3$. Our model has more couplings and the ``mWTI surface'' is a higher-dimensional sub-manifold.}
\label{fig:mWTI_surface}
\end{figure}

We have, then, flow equations for $\Aomega,Y,\zeta,\kappa,\delta,g$ and $\lambda$, and mWTIs constraining $\delta,g$ and $\lambda$. To implement the latter, we shall follow the standard approach from the high-energy literature~\cite{HG2012,HGJJCW2004} of treating $\Aomega,Y,\zeta$ and $\kappa$ as ``independent'' variables and computing them from their flow equations, then fixing the remaining couplings using their mWTIs alone. This tends to sum a larger class of diagrams than one would in computing the dependent couplings from their flow equations~\cite{HGJJCW2004}, and so presumably gives a more robust description of criticality. In terms of the RG flow lines in the space spanned by the couplings, this constraining corresponds to forcing trajectories to lie in the sub-manifold defined by the mWTIs, as depicted in Fig.~\ref{fig:mWTI_surface}. There, the endpoint of the constrained trajectory is more likely to finish close to the true low-energy effective average action; an unconstrained flow runs the risk of being driven away from the sub-manifold by an RG-relevant coupling.

\section{Results}
\label{sec:results}

In addition to solving the system in the presence of mWTI constraints, we also compute flows and fixed-point properties using the flow equations for all seven couplings, ignoring the mWTIs, and compare the results to those obtained with mWTIs enforced. This should give insight into the effect on the modelling of rigorously observing symmetry constraints. We shall refer to the flows in which mWTIs are enforced as ``constrained'' and the others as ``unconstrained''.

Both the constrained and unconstrained flows feature a NFL fixed point. The fixed-point couplings and anomalous dimensions depend on the non-universal parameter $N=k_F/\kUV$. The appearance of $k_F$ in the low-energy theory is an acceptable, ``weak'' non-universality --- indeed, such UV-IR mixing is a well-established phenomenon in theories featuring a Fermi surface coupled to a gapless boson~\cite{IMSL2015,AFGTHW2015}. 't Hooft anomaly preservation~\cite{DERTTS2021} also requires this dependence to be retained along the flow. However, dependence of the universal behaviour on $\kUV$, the size of the region about the Fermi surface in which we linearize the dispersion, is {\it prima facie\/} not admissible. The size of this region depends strongly on the details of the initial dispersion, i.e.\ on the microscopic hopping parameters, which is highly non-universal information. Thus, the $N$-dependence of the critical data is in general problematic.

However, we argue that the limit $N\to \infty$ is physically relevant. For one, this mimics what occurs in a momentum-cutoff scheme whereby the momentum cutoff, i.e.\ the width of the annulus of momenta considered about the Fermi surface, is sent to zero. More telling, perhaps, is the fact that only this limit reproduces the loop-U(1) symmetry that most NFLs are claimed to have~\cite{DERTTS2021,DETS2021,XHALUM2024}. At any finite $N$, our low-energy theory (if we were to reinclude four-fermion interactions) would not have a loop-U(1) symmetry, because scattering events in which the incoming momenta are changed by an angle of order $1/N$ would be allowed. The emergent symmetry and its associated anomaly are generically important for correctly capturing the low-energy physics~\cite{DERTTS2021,DETS2021,XHALUM2024}, so we must take $N\to\infty$. Furthermore, the term $g^2 f(\delta,\zeta,Y)$ in the mWTI~(\ref{eqn:mWTIs}) does not go to zero at the end of the flow, as it ought to in order to restore gauge symmetry, but since it is weighted by a factor of $1/N$ this effect vanishes in the $N\to\infty$ limit. We interpret the $N$-dependence as an artefact of the frequency-cutoff scheme --- in any scheme in which momenta also appeared in the cutoff function, $\kUV$ would presumably flow to zero as $\Lambda \to 0$. Thus, though the following results are of some interest at finite $N$, they are most trustworthy for $N\to \infty$.

\setlength{\tabcolsep}{20pt}
\renewcommand{\arraystretch}{1.5}
\begin{table*}[t]
\centering
\caption{Fixed-point values of the fermion anomalous dimensions, boson dynamical exponent and couplings. $\ell(N)$ is $(\log N)/N$ and the ellipses denote $\mathcal{O}(1/N)$ terms.}
\begin{threeparttable}
\begin{ruledtabular}
\begin{tabular}{c c c}
& Unconstrained & Constrained \\
\hline
$\eta_\omega$ & $\frac{1}{2}-0.26~\ell(N)+\dots$ & $\frac{1}{2}-0.47~\ell(N)+\dots$\\
$\eta_\mathbf{k}$ & $-0.21~\ell(N)+\dots$ & $-0.47~\ell(N)+\dots$ \\
$z_\mathbf{A}$ & $2-0.19~\ell(N)+\dots$& 2 \\
$\zeta$ & $\frac{0.31}{N}\left[1+0.43~\ell(N)+\dots\right]$ & $\frac{0.86}{N}\left[1+0.34~\ell(N)+\dots\right]$\\
$\kappa$ & $\infty$~(\dag) & $\sqrt{2}-0.44~\ell(N)+\dots$ \\
$\delta$ & $12-3.4~\ell(N)+\dots$& $4-1.7~\ell(N)+\dots$ \\
$g$ & $\sqrt{2}-0.14~\ell(N)+\dots$& $\sqrt{2}-0.17~\ell(N)+\dots$\\
$\lambda$ & $4~\ell(N)+\dots$ & 0
\end{tabular}
\end{ruledtabular}
\begin{tablenotes}
\item[$\dag$] Strictly this result holds at large but not infinite $N$.  At $N=\infty$ in the unconstrained case, $\kappa$ flows to some finite constant (generally not $\sqrt{2}$) that depends strongly on initial conditions. 
\end{tablenotes}
\end{threeparttable}
\label{tab:fp_vals}
\end{table*}

\begin{figure}[h]
\vspace{4mm}
\centering
\includegraphics[width=0.8\columnwidth]{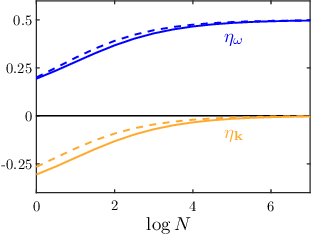}
\caption{The fermion anomalous dimensions as functions of the non-universal parameter $\log N$ ($N=k_F/\kUV$). The solid lines are from the ``constrained'' analysis (i.e.\ with modified Ward-Takahashi identities enforced); the dashed lines are from the unconstrained analysis.}
\label{fig:collate_etas}
\end{figure}

The fixed point couplings and exponents have been calculated numerically at all $N$, and their large-$N$ limits are given in Table~\ref{tab:fp_vals}. There, we use the expression 
\begin{equation}
\label{eqn:zA}z_\mathbf{A}=\frac{2}{2-\etaAomega+\etaAq}
\end{equation}
for the dynamical exponent, which differs from other expressions employed~\cite{SMPS2016,MTCH2018} but agrees with them at $\etaAq=0,\etaAomega\approx 1$, which obtains here. 
In addition, the values at all $N$ are displayed for $\eta_\omega$ and $\eta_\mathbf{k}$ in Fig.~\ref{fig:collate_etas} and for $\lambda$ in Fig.~\ref{fig:lambda_unconstr}. The fact that $\eta_\mathbf{k}\to0$ as $N\to\infty$ is consistent with previous works on U(1) gauge field criticality~\cite{SCRNOS1995,JP1994} (though these did start with a Hertz-Millis-like, $\propto |\Omega|/|\mathbf{q}|$ term in the boson propagator).  

\subsection{Effect of mWTI constraints}
\label{subsec:results_constraints}
It is apparent that, though the constrained analysis produces some differences in the fixed point values, most couplings have the same value in the $N\to\infty$ limit. It is particularly remarkable that, in so simple a truncation, the same large-$N$ value of $z_\mathbf{A}$ arises. Though one cannot be sure as to cause and effect in solutions of equations, this seems be a consequence of genuinely different mechanisms in the constrained and unconstrained cases --- in the former mostly due to mWTIs, and in the latter to scaling relations. Note, however, that the reason that $\eta_\omega\to 1/2$ and $\eta_\mathbf{k} \to 0$ is common to both. A complete analysis is given in Appendix~\ref{append:z=2}. 

It is difficult to draw strong conclusions from this --- the most obvious one would be that enforcing mWTIs is not necessary in order to describe most of the properties of the fixed point (though some values are affected, as well as the topology of the flow about criticality --- see below). However, one might also speculate that Table~\ref{tab:fp_vals} is evidence that the soft-frequency-cutoff scheme is particularly good at accessing the strong-coupling régime. This is of course completely unsubstantiated until an analysis of the system is performed with, say, a soft momentum cutoff, which is beyond the scope of this work.

\begin{figure}
\vspace{4mm}
\centering
\includegraphics[width=0.8\columnwidth]{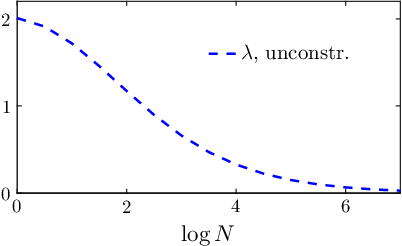}
\caption{The fixed-point value of the four-point coupling as a function of $\log N$ ($N=k_F/\kUV$), in the unconstrained case. It is identically 0 in the constrained case.}
\label{fig:lambda_unconstr}
\end{figure}

\begin{figure}
\centering
\subcaptionbox*{}{\includegraphics[width=0.8\linewidth]{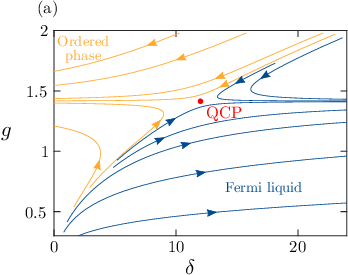}}
\subcaptionbox*{}{\hspace{1mm}\includegraphics[width=0.79\linewidth]{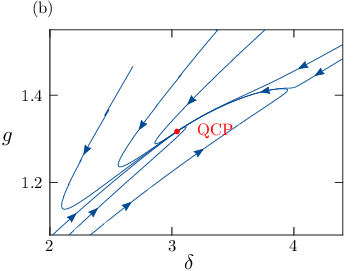}}
\caption{The functional renormalization group flow (a) without and (b) with modified Ward-Takahashi identities enforced. We have taken $N\to\infty$ in (a), and $N=1$ in (b) (this is to facilitate easy plotting; the flow topology is insensitive to $N$). The flows in (a) have all the other couplings set at their fixed-point values, while those in (b) have been projected into the $(\delta,g)$ plane. The regions in (b) with no flow lines correspond to points that could not satisfy the modified Ward-Takahashi identities for any values of the other couplings (see the main text).}
\label{fig:flow_topo}
\end{figure}

There are more considerable differences between the two approaches in the behaviour of some of the other couplings. For example, the four-point coupling $\lambda$ is identically zero at the fixed point in the constrained case (as required by restoration of gauge symmetry at the end of the flow), while it only tends to zero as $N\to\infty$, as in Fig.~\ref{fig:lambda_unconstr}, in the unconstrained case. More significant still is the behaviour of the mass term $\delta$ in the two cases. The large-$N$ mass obtained in the unconstrained case, 12, is almost such that perturbative expansions could be performed in its inverse; the constrained mass 4, however, is more emphatically in the $\mathcal{O}(1)$ coupling régime. Aside from the differing fixed-point values, the behaviour of the flow about criticality is also very different, in that we find $\delta$ a relevant operator at the fixed point in the unconstrained case (as one would expect for the criticality of a standard symmetry-breaking transition) but irrelevant in the constrained case. This is illustrated in Fig.~\ref{fig:flow_topo}, where we have shown constrained and unconstrained flows about the fixed point in the $(\delta,g)$ plane. In the unconstrained flow in Fig.~\ref{fig:flow_topo}(a), all the other couplings have been set to their fixed-point values, while in the constrained flow in Fig.~\ref{fig:flow_topo}(b) various initial conditions for $\kappa$ and $\zeta$ have been used and the flow projected to the $(\delta,g)$ plane. This was necessary because $\delta$ and $g$ are fixed by the other couplings through the mWTIs. Furthermore, it was found that many initial values of $\delta$ and $g$ could not be used, as they couldn't satisfy the mWTIs for any choice of the other couplings --- hence the large regions in Fig.~\ref{fig:flow_topo} (b) that have no flow lines. Graphically, this means that the sub-manifold of coupling space parametrized by the mWTIs must intersect the majority of the $(\delta,g)$ planes almost perpendicularly. 

In the unconstrained case, then, there is an instability towards a phase with $\delta<0$ (the phase $\delta\to\infty$ is just the normal Fermi liquid). This description of the ordering transition, though appropriate to actual symmetry-breaking transitions, is incorrect here as it would result in an expectation value for the gauge field, and a gauge symmetry cannot be spontaneously broken (as is well known~\cite{SE1975}). There is an ordered phase, indeed, but it can presumably only be rigorously understood in terms of restoring of a higher-form symmetry~\cite{DGAKNS2015,PG2023}. The mWTIs correctly ensure that the gauge-field mass is irrelevant about the critical point, and so the constrained case gives a more faithful description of the flow topology.

A final (if somewhat inconsequential) observation: the large-$N$ value $\kappa\sim\sqrt{2}$ in the constrained case gives us the full scaling form of a coupling, \emph{viz.}
\begin{equation}
\BAomega\sim \frac{N e^2}{16\pi\Lambda}~.
\end{equation}
The coefficient of such a power law is rarely available from the RG.

\subsection{Comparison with the literature}
\label{subsec:results_compare}

Let us now compare our large-$N$ results with previous works on U(1)-gauge-field-induced criticality. In Table~\ref{tab:comparison} we summarize key aspects of the critical behaviour, as calculated in this work and previous studies (there, $\Sigma$ is the fermion self-energy). The ``Hertz-Millis-type'' works model Landau damping by using a boson propagator including the one-loop correction given by the particle-hole bubble, and then finding $\Sigma(\omega,k_F)$ from one-loop perturbation theory with the dressed boson propagator (this is equivalent to the results of the RPA). Refs.~\cite{THWM2015,MMSS2010a} go further and calculate loop corrections to this RPA form, finding that $z_\mathbf{A}$ increases above 3 and the small-$\omega$ power-law exponent of $\Sigma(\omega,k_F)$ increases above 2/3. Ref.~\cite{DKPS1993} also uses a Hertz-Millis-like boson propagator, but then finds the fermionic properties by an approximate eikonal solution of the Dyson equation, resulting in exponential behaviour for $\Sigma(\omega,k_F)$ at small $\omega$. The eikonal expansion is an improvement on one-loop perturbation theory and even yields some non-perturbative results, but the method still does not treat all IR divergences on the same footing.

We also note an early work~\cite{CNFW1994} that studies finite-density fermions coupled to a Chern-Simons U(1) gauge field. There is also a four-fermion, $1/|\mathbf{q}|^x$ interaction, and the authors find a controlled expansion in $(1-x)$. Extrapolating to $x=0$ gives $\Sigma(\omega,k_F)\propto\omega^{1/2}$, and furthermore enforcing the equation of motion of $A_0$ (which is a constraint) transforms the four-fermion interaction into a Maxwell term for $\mathbf{A}$. These results are tantalizing, but it is difficult to compare them with our own since (a) the gauge field has no dynamics in that work, and (b) both the $\mathcal{T}$ and $\mathcal{P}$ symmetries are broken by the Chern-Simons term. 

\setlength{\tabcolsep}{6pt} 
\begin{table}[h]
\centering
\caption{The boson dynamical exponent, and fermion self energy at the Fermi surface, for this study and some previous works. $C$ in the penultimate row is a positive constant. Note that in our scheme $\Sigma(\omega,k_F)\propto\omega^{1-\eta_\omega}$.}
\begin{ruledtabular}
\begin{tabular}{c c c}
& $z_\mathbf{A}$&$\Sigma(\omega,k_F) \propto$\\
\hline
This work & 2 & $\omega^{1/2}$\\
\begin{minipage}{4cm} \vspace*{1.5mm}Hertz-Millis-type\\\cite{JP1994,YKAFXW1994,BHPLNR1993,BBHM1993,PL1989,BALIAM1994}\end{minipage} & 3 & $\omega^{2/3}$\\
$\vphantom{2^\dagger}$H-M-type, $\geq 3$-loop~\cite{THWM2015,MMSS2010a} & $3.02$ & $\omega^{0.68}$ \\
H-M/eikonal expansion~\cite{DKPS1993} & 3 & $\omega^{5/4}e^{C/\sqrt{\omega}}$\\
$(1-x)$ - expansion~\cite{CNFW1994} & $-$ & $\omega^{1/2}$
\end{tabular}
\end{ruledtabular}
\label{tab:comparison}
\end{table}

Our results of 2 and 1/2, respectively, are thus largely at odds with the literature. However, we argue that our results are plausible, in particular in light of some studies in recent years that indicate that many NFLs induced by $\mathbf{q}=\mathbf{0}$ gapless bosons in $d=2$ also have $z=2$. None of these consider the case of a U(1) gauge field, and so are not directly comparable; nevertheless, they at least demonstrate that dynamical exponents $z$ significantly less than 3 are possible for NFLs associated to $\mathbf{q}=\mathbf{0}$ transitions, contrary to the more widely held view. It should be mentioned that one key difference in our work is that the gauge field Yukawa coupling, being $\propto \hat{\mathbf{k}}$, couples with opposite signs to fermions on opposite sides of the Fermi surface, which is not true of any of the below. Ref.~\cite{YSSLSK2016} studies the Ising-nematic transition by quantum Monte Carlo and finds $z\approx2$, though very little departure of the fermionic exponents from the Fermi-liquid form; Ref.~\cite{YLWJAK2022}, on the other hand (also Monte Carlo), considers XY rotors interacting with itinerant spins near a ferromagnetic critical point and finds $z=2$, $\Sigma(\omega,k_F)\propto \omega^{1/2}$. However, in both of these cases the order parameter is not conserved, which is untrue of the gauge-field case, and in fact Ref.~\cite{YLWJAK2022} invokes this fact as an explanation for the $z=2$ dynamics. More encouragingly, a recent work~\cite{MHPJHY2024} considers a conserved order parameter which couples to the fermion density, and also finds $z\approx2$. The authors construct a generalized Hertz-Millis approach in which the fermions are integrated out sequentially, using a momentum cutoff. 

\section{Conclusions}
\label{sec:conclusion}

In this paper we have presented a functional renormalization group (fRG) analysis of the quantum critical point hosted by a U(1) gauge field coupled to fermions with a circular Fermi surface (in $d=2$), and examined the effect of constraints imposed by Ward-Takahashi identities (WTIs). The non-Fermi-liquid (NFL) state at this critical point is of interest in its own right and has not received a non-perturbative RG treatment in the past; in addition, the constraints dictated by gauge symmetry might act to improve the design of new modelling schemes for NFLs, or test the soundness of old ones. 

Our fRG scheme utilizes a soft frequency cutoff function~\cite{SMPS2016,MTCH2018}, which allows Landau damping of the gauge field to develop gradually as the frequency scale $\Lambda$ is lowered, without the need to integrate over all fermionic modes at the start of the flow; our effective average action is thus local at intermediate scales. This regularization breaks the gauge symmetry, resulting in modified WTIs (mWTIs), which act to constrain the RG flow in parameter space. We do not treat instanton effects, assuming that the associated additions to the effective action would be RG-irrelevant in the presence of a Fermi surface~\cite{SL2008,KK2005}. The question of whether such terms should be included for compact U(1) gauge theory, and whether they could force the theory to always be confining, is an open one~\cite{YZKLYW2012,IHBSSS2003,SL2008,KK2005,GAIITM2006} --- as such our conclusions may or may not be pertinent to compact U(1), but are certainly applicable to non-compact U(1) gauge theory.

We find a NFL fixed point, which depends on the non-universal ratio $N=k_F/\kUV$, though we argue that this dependence is an artefact and the $N\to\infty$ limits are most trustworthy. We find that enforcing mWTIs does not change many of the fixed-point values of the couplings, which may indicate that such constraints are superfluous, or that the chosen frequency-cutoff scheme is particularly good at describing NFLs. The behaviour of the boson mass term does vary, however. Without mWTI constraints it has to be tuned in order to reach criticality, else the flow may go to $\delta<0$ (which is forbidden by gauge symmetry~\cite{SE1975}), while enforcing the mWTIs makes it irrelevant about the fixed point. This (correctly) signifies that an analysis by higher-form symmetries is necessary to describe the ordered phase. 

One might wonder whether there exist other models in which constraints set by mWTIs induce a larger difference in the fixed-point properties. A natural extension would be to an SU($N_c$) gauge field with more fermion flavours, since then there are more identities with different group structure to the Abelian ones. The $N_c=3$ case would of course be particularly relevant to dense quark-gluon plasmas. Another possible line of enquiry would be to consider ``larger'', emergent symmetry groups, such as the loop-U(1) symmetry enjoyed by IR theories without non-forward scattering~\cite{DERTTS2021}. Since the latter is an infinite-dimensional Lie group, formulating its WTIs could prove very difficult, however.

Our results for the critical exponents, most notably $z_\mathbf{A}=2$ for the gauge field dynamical exponent and $\Sigma(\omega,k_F)\propto\omega^{1/2}$, differ distinctly from the accepted $z_\mathbf{A}=3$, $\Sigma(\omega,k_F)\propto\omega^{2/3}$~\cite{JP1994,YKAFXW1994,BHPLNR1993,BBHM1993,PL1989,BALIAM1994,MMSS2010a,THWM2015} --- however, our scheme takes account of Landau damping in a consistent, local way, and does not fully incorporate the effects of IR fermionic modes until the end of the flow. We have also argued that our results are quite plausible in light of various works in recent years that find $z=2$ for other universality classes with a critical $\mathbf{q}=\mathbf{0}$ boson. A quantum Monte Carlo study of the U($1$)-gauge-field-induced NFL would pronounce more definitively on the topic.

\section*{Acknowledgments}

The authors would like to thank B. Braunecker and Z. D. Shi for helpful discussions. TPS acknowledges funding from the EPSRC (UK) under grant EP/T518062/1. CAH is grateful for the hospitality of the Max Planck Institute for the Physics of Complex Systems (MPI-PKS) in Dresden, Germany, where part of this work was carried out.


\begin{appendices}
\makeatletter\renewcommand*{\@seccntformat}[1]{
\MakeUppercase{\appendixname} \thesection:
}\makeatother

\makeatletter
\renewcommand{\theequation}{\thesection\arabic{equation}}
\renewcommand{\thefigure}{\thesection\arabic{figure}}
\renewcommand{\bibnumfmt}[1]{[\thesection#1]}

\setcounter{equation}{0}
\setcounter{figure}{0}
\setcounter{table}{0}
\section{Functions used in main text}
\label{append:integrals}
We collate here the integral expressions for the various functions appearing in the mWTIs and flow equations. In all cases, we also give a small-$\zeta$ expansion of the integral, with $Y\to \infty$. Note that these expansions begin with a $\propto 1/\zeta$ term, then a $\propto \log \zeta$ term (both of these may have coefficient 0), and a subsequent Maclaurin-style series in $\zeta$.

First, there is that in the mWTI for $g$:
\begin{widetext}
\begin{align}
f(\delta,\zeta,Y)&=\frac{4}{\pi}\int_0^\infty da~\frac{a^2}{(a^2+1)^2}\int_0^Y \frac{d\rho}{\rho}\frac{1}{a^2+1+\zeta^2\rho^2+\delta}\left(1-\frac{a}{\sqrt{a^2+\rho^2}}\right)\nonumber\\&=-\frac{\log \zeta}{\left(\sqrt{\delta+1}+1\right)^2}+\mathcal{O}(\zeta^0)~.
\end{align}
The functions $F_i$ appearing in $\eta_\omega$ and $\eta_\mathbf{k}$ are given by
\allowdisplaybreaks
\begin{align}
\label{eqn:F_fns}
F_1(\delta,\zeta,Y)&=i\int_0^\infty \frac{da}{2\pi^2}~\frac{a^3}{a^2+1}\int_0^Y\frac{d\rho}{\rho}\frac{1}{(a^2+1+\zeta^2\rho^2+\delta)^3}(a-\sqrt{a^2+\rho^2})\nonumber\\
&= -\frac{3i}{32\pi\zeta}~g_1(\delta)-\frac{i\log\zeta}{32\pi}\frac{\sqrt{\delta+1}+3}{\sqrt{\delta+1}\left(\sqrt{\delta+1}+1\right)^3}+\mathcal{O}\left(\zeta^0\right)~,\nonumber\\
F_2(\delta,\zeta,Y) &= \int_0^\infty \frac{da}{2\pi^2}~\frac{a^3}{(a^2+1)^2}\int_0^Y d\rho~ \frac{1}{(a^2+1+\zeta^2\rho^2+\delta)^2}\frac{a}{\rho}\left(-1+\frac{a}{\sqrt{a^2+\rho^2}}\right)\nonumber\\
&= \frac{1}{8\pi}\frac{\log\zeta}{\left(\sqrt{\delta+1}+1\right)^3}+\mathcal{O}\left(\zeta^0\right)~,\nonumber\\
F_3(\delta,\zeta,Y)&=\int_0^\infty \frac{da}{2\pi^2}~\frac{a^3}{(a^2+1)^2}\int_0^Y d\rho~ \frac{1}{(a^2+1+\zeta^2\rho^2+\delta)^2}\frac{i\left(1+2(a/\rho)^2-2(a/\rho)\sqrt{1+(a/\rho)^2}\right)}{\sqrt{1+(a/\rho)^2}}\nonumber\\
&= \frac{i}{8\pi\zeta}~g_2(\delta)+\frac{i\log\zeta}{4\pi}\frac{1}{\left(\sqrt{\delta+1}+1\right)^3}+\mathcal{O}\left(\zeta^0\right)~,\nonumber\\
F_4(\delta,\zeta,Y)&= -\int_0^\infty \frac{da}{4\pi^2} ~\frac{a^2}{a^2+1}\int_0^Y \frac{d\rho}{\rho} \frac{a^2+1+3\zeta^2\rho^2+\delta}{(a^2+1+\zeta^2\rho^2+\delta)^3}\left(1+2(a/\rho)^2-2(a/\rho)\sqrt{1+(a/\rho)^2}\right)\nonumber\\
&=\frac{\log\zeta}{16\pi}\frac{1}{\sqrt{\delta+1}\left(\sqrt{\delta+1}+1\right)^2}+\mathcal{O}\left(\zeta^0\right)~,\nonumber\\
F_5(\delta,\zeta,Y)&= \int_0^\infty \frac{da}{2\pi^2}~ \frac{a^2}{(a^2+1)^2}\int_0^Y \frac{d\rho}{\rho} \frac{a^2+1+2\zeta^2\rho^2+\delta}{(a^2+1+\zeta^2\rho^2+\delta)^2}\frac{i a \left(1+2(a/\rho)^2-2(a/\rho)\sqrt{1+(a/\rho)^2}\right)}{\sqrt{\rho^2+a^2}}\nonumber\\
&=\frac{i}{16\pi}\frac{1}{\left(\sqrt{\delta+1}+1\right)^2}+\mathcal{O}\left(\zeta\right)~,\nonumber\\
F_6(\delta,\zeta,Y)&=\int_0^\infty \frac{da}{4\pi^2}~ \frac{a^2}{(a^2+1)^2}\int_0^Y \frac{d\rho}{\rho} \frac{a^2+1+2\zeta^2\rho^2+\delta}{(a^2+1+\zeta^2\rho^2+\delta)^2}\left(1+6(a/\rho)^2-\frac{2 a (2+3(a/\rho)^2)}{\sqrt{a^2+\rho^2}}\right)\nonumber\\
&=-\frac{\log\zeta}{16\pi}\frac{1}{\left(\sqrt{\delta+1}+1\right)^2}+\mathcal{O}\left(\zeta^0\right)~,\nonumber\\
\end{align}
where 
\begin{gather*}
g_1(\delta)=\frac{(\delta+1)\left[3\sqrt{\delta}\log\left(2\delta+1-2\sqrt{\delta+1}\sqrt{\delta}\right)+2\sqrt{\delta+1}\right]+2(\delta+1)^{5/2}-4\sqrt{\delta+1}}{6\delta^3(\delta+1)}~,\nonumber\\
g_2(\delta) = -\frac{3\sqrt{\delta+1}}{2\delta^2}-\frac{(2\delta+3)\log\left(2\delta+1-2\sqrt{\delta}\sqrt{\delta+1}\right)}{4\delta^{5/2}}~.
\end{gather*}
Finally, the functions $G_i$ in $\eta_{\tilde{g}}$ are given by
\begin{align}
G_1(\delta,\zeta,Y)&= \int_0^\infty \frac{da}{2\pi^2}~\left(\frac{a^2}{a^2+1}\right)^2\int_0^Y\frac{d\rho}{\rho}\frac{1}{(a^2+1+\zeta^2\rho^2+\delta)^2}\left(-1+\frac{a}{\sqrt{\rho^2+a^2}}\right) \nonumber\\
&= \frac{\log\zeta}{8\pi}\frac{1}{\left(\sqrt{\delta+1}+1\right)^3}+\mathcal{O}\left(\zeta^0\right) ~,\nonumber\\
G_2(\delta,\zeta,Y)&=\int_0^\infty \frac{da}{4\pi^2}~\frac{a^4}{(a^2+1)^3}\int_0^Y \frac{d\rho}{\rho}\frac{1}{a^2+1+\zeta^2\rho^2+\delta}\frac{i a/\rho}{(1+(a/\rho)^2)^{3/2}}\nonumber\\
&= \frac{i}{64\pi}\frac{3\sqrt{\delta+1}+1}{\left(\sqrt{\delta+1}+1\right)^3}+\mathcal{O}\left(\zeta\right)~,\nonumber\\
G_3(\delta,\zeta,Y)&= \int_0^\infty \frac{da}{4\pi^2}~\frac{a^4}{(a^2+1)^3}\int_0^Y \frac{d\rho}{\rho}\frac{1}{a^2+1+\zeta^2\rho^2+\delta}\left[2-\frac{(a/\rho)(3+2(a/\rho)^2)}{(1+(a/\rho)^2)^{3/2}}\right]\nonumber\\
&= -\frac{\log\zeta}{32\pi}\frac{3\sqrt{\delta+1}+1}{\left(\sqrt{\delta+1}+1\right)^3}+\mathcal{O}\left(\zeta^0\right) ~.
\end{align}
\end{widetext}

\setcounter{equation}{0}
\setcounter{figure}{0}
\setcounter{table}{0}
\section{Mechanisms for $z_\mathbf{A}\to 2$ at large $N$}
\label{append:z=2}

Here, we briefly summarize the arguments leading to $z_\mathbf{A}=2$ in the large-$N$ limit for both the constrained and unconstrained cases, in order to assess whether they genuinely arrive at the same result through different ``mechanisms''. Common to both is the fact that $\etaAomega = C g^2$, for a positive constant $C$ of order 1. This is a consequence of the frequency-cutoff scheme: though the diagram generating this term, the particle-hole bubble (see Appendix~\ref{append:deriv}), will always be proportional to $\tilde{g}_\mathbf{A}^2$, which of the other RG variables ($\Aomega$, etc.) also appear in the expression (and consequently in the definition of $g$) is scheme-dependent. As such, the effect this has on the value of $z_\mathbf{A}$ is difficult to isolate. Common too to both is the fact that $\etaAq=0$ at criticality.

Then, in the constrained case, the mWTI further gives $g\propto\kappa$ (up to a positive constant, which ideally ought to be exactly 1), so that if there is to be Landau damping at criticality (i.e.\ $\etaAomega\neq 0$) we must have $\kappa\sim$ const., which is equivalent to $\etaAomega=1$, so from~(\ref{eqn:zA}) we have $z_\mathbf{A}=2$.

In the unconstrained case, $\etaAomega= Cg^2$ and the presence of Landau damping give $g\sim$ const., so that
\begin{equation}
\label{eqn:g_scaling_rel}
\eta_{\tilde{g}}-\eta_\mathbf{k}+\frac{1}{2}-\frac{1}{2}\etaAomega=0~.
\end{equation}
Then: overdamping of the boson suggests $\zeta\to 0$ or a small constant. The former is not possible, since then $\eta_\omega$ tends to blow up (as it is $\mathcal{O}(1/\zeta)$ (see Eqns.~(\ref{eqn:anom}) and~(\ref{eqn:F_fns}))), so we must have
\begin{equation}
\frac{\beta_\zeta}{\zeta}=\eta_\mathbf{k}-\eta_\omega+\frac{\etaAomega}{2}=0~.
\end{equation}
Added to Eqn.~(\ref{eqn:g_scaling_rel}), this gives
\begin{equation}
\eta_\omega-\eta_{\tilde{g}}=\frac{1}{2}~.
\end{equation}
The left hand side here $\sim (1/N)\times \mathcal{O}(1/\zeta)$ at small $\zeta$, so $\zeta=\mathcal{O}(1/N)$. Then, $\eta_\mathbf{k}$ and $\eta_{\tilde{g}}$ go to zero as $N\to\infty$, because they both $\sim(1/N)\times \mathcal{O}(\log\zeta)$. Thus, $\etaAomega\to 1$ from~(\ref{eqn:g_scaling_rel}), and $z_\mathbf{A}\to 2$, as $N\to\infty$.

\setcounter{equation}{0}
\setcounter{figure}{0}
\setcounter{table}{0}
\section{Possible further mass terms for the Coulomb field}
\label{append:nonpert}

Here we provide some discussion on the possibility of a gauge-invariant mass term for the $\phi$ field, as discussed below~(\ref{eqn:Pi_param}). Mass terms for $\mathbf{A}$ and $\phi$ would entail adding constant terms to the ansatzes for $\Pi_\mathbf{A}$ and $\Pi_\phi$:
\begin{align}
    &\Pi_\mathbf{A}(q)=(\BAomega-1)\Omega^2+(\BAq-1)|\mathbf{q}|^2+\mathcal{B}_\mathbf{A}~,\nonumber\\
    &\Pi_\phi(q)=(\Bsq-1)|\mathbf{q}|^2+\mathcal{B}_\phi~.
\end{align}
These additions respect gauge symmetry, contrary to the standard intuition. However, whether or not they arise in a given scenario is a delicate (and gauge-dependent) question~\cite{MPDS1995}: they give rise to non-analyticities in $\Pi^{\mu\nu}$, but on the other hand are known to occur in some scenarios. Indeed, the Thomas-Fermi screening wavevector for $\phi$ is just such a mass term. Thus, even if one were to neglect $\mathcal{B}_\mathbf{A}$ on grounds of non-analyticity, one might still need to retain $\mathcal{B}_\phi$ on the grounds that $\phi$ acts non-locally in the Coulomb gauge~\cite{OBBG1967} (though the transverse part of $\mathbf{A}$ does not), and so perhaps its part of the effective action might have non-local terms.  In our calculation this question is rendered unimportant because the $\phi$ field is screened out and can therefore be dropped.

The above discussion is related to the artefact of the frequency scheme mentioned in the main text, whereby when $M_\phi^2$ is constrained by its mWTI~(\ref{eqn:mWTI_mass}), it does not vanish as it should at the end of the flow, when gauge symmetry is recovered. Ignoring the corrections to the mWTI for $M_\phi^2$ in (\ref{eqn:brief_Yukawa_mWTI}), we have $\tilde{g}_s=e\Aomega$, and substituting this into the first mWTI in~(\ref{eqn:mWTI_mass}) gives $M_\phi^2=k_{TF}^2$, the square of the Thomas-Fermi screening wavevector evaluated with flowing couplings. This is somewhat suggestive: the fact that $M_\phi^2$ does not vanish as $\Lambda \to 0$ is definitely an undesirable artefact, but it should also be noted that the term $M_\phi^2\delta^\mu_0\delta^\nu_0$ in $\Pi^{\mu\nu}$ is not purely longitudinal, and so may be acting as a partial proxy for a transverse mass term.

\begin{widetext}
\setcounter{equation}{0}
\setcounter{figure}{0}
\setcounter{table}{0}
\section{Derivations of Flow Equations and mWTIs}
\label{append:deriv}

In this appendix we sketch the derivation of selected flow equations and mWTIs, starting from the Wetterich equation~(\ref{eqn:Wetterich}) and master mWTI~(\ref{eqn:mastermWTI}), respectively. 

Consider the Wetterich equation, first. We proceed by substituting the ansatz~(\ref{eqn:Ansatz}) into it, and then comparing the coefficients of monomials in $\mathbf{A}$ and $\psibar\psi$ (arguments suppressed) on the left- and right-hand sides. The inverse inside the supertrace first requires some attention, however. We can write~\cite{WMMSCH2012} (again dropping superscripts $\Lambda$) 
\begin{equation}
\mathbf{\Gamma}^{(2)}[\psi,\psibar,\mathbf{A}]+\mathbf{R} = \mathbf{G}^{-1}-\tilde{\mathbf{\Sigma}}[\psi,\psibar,\mathbf{A}]~,
\end{equation}
where $\mathbf{G} = \text{diag}(G,-G^T,D)$ is the matrix of dressed propagators, and $\tilde{\mathbf{\Sigma}}$ contains all terms in the above that are non-constant in the fields (i.e.\ the interaction vertices). The supertrace term then becomes
\begin{equation}
\text{Str}\left[\mathbf{G}(\partial_\Lambda\mathbf{R})\left(\openone-\mathbf{G}\tilde{\mathbf{\Sigma}}\right)^{-1}\right]~,
\end{equation}
and the inverse can be expanded in a Neumann series. Coefficients of field monomials can then be matched. 

In the following, we shall explicitly evaluate loop integrals in some select flow equations and mWTIs. We shall need the so-called ``single-scale propagators'':

\begin{align}
S_f(k)=\partial_\Lambda^R G(k) = -\frac{\omega^2\Lambda}{(\omega^2+\Lambda^2)^2}\frac{i(2-\eta_\omega)\Aomega\omega-(2-\eta_\mathbf{k})\Ak l}{(i\Aomega\omega-\Ak l)^2}~,
\end{align}
and
\begin{align}
S_\mathbf{A}^{ij}(q)=\partial_\Lambda^RD^{ij}(q)=\frac{\BAomega\Lambda(2-\etaAomega)}{\left(\BAomega(\Omega^2+\Lambda^2)+\BAq|\mathbf{q}|^2+M_\mathbf{A}^2\right)^2}\left(g^{ij}+\frac{q^iq^j}{\mathbf{q}^2}\right)~,
\end{align}
where $\partial_\Lambda^R=\sum_{i=b,f}\partial_\Lambda R_i\frac{\partial}{\partial R_i}$.
\\\vspace{2mm}
\begin{center}
\textbf{1. Flow equations with purely bosonic or fermionic loops}
\end{center}

The flows for $\BAomega,\BAq,M_\mathbf{A}^2$ and $\tilde{\lambda}$ fall into this category. The flow equation for $\BAomega$ is 
\begin{align}
\partial_\Lambda\BAomega=\tilde{g}_\mathbf{A}^2~\frac{\partial^2}{\partial\Omega^2}\biggr\rvert_{q=0}\left[\partial_\Lambda^R\int_k G(k)G(k+q) \sin^2\theta\right]~,
\end{align}
where $\theta$ is the angle between $\mathbf{k}$ and $\mathbf{q}$. The flow equation for $\BAq$ is
\begin{align}
\partial_\Lambda\BAq=-\frac{\tilde{g}_\mathbf{A}^2}{2} ~g^{ij}~\frac{\partial^2}{\partial q^i\partial q^j}\biggr\rvert_{q=0}\left[\partial_\Lambda^R\int_k G(k)G(k+q) \sin^2\theta\right]~.
\end{align}
The flow equation for $M_\mathbf{A}^2$ is
\begin{align}
\label{eqn:fullflow_M}
\partial_\Lambda\left(M_\mathbf{A}^2\right)=2\tilde{g}_\mathbf{A}^2~\partial_\Lambda^R\int_k G(k)^2 \sin^2\theta-\tilde{\lambda}~\partial_\Lambda^R\int_{q'}g^{ij}D_{ij}(q')~.
\end{align}
The flow equation for $\tilde{\lambda}$ is
\begin{align}
\label{eqn:fullflow_lambda}
\partial_\Lambda\tilde{\lambda}=\frac{3}{2}~\tilde{g}_\mathbf{A}^4~ \partial_\Lambda^R\int_k G(k)^4-\frac{9}{4}~\tilde{\lambda}^2~\partial_\Lambda^R\int_q D_{ij}(q)D^{ij}(q)~.
\end{align}
As an example of a purely fermionic loop integral, let us evaluate the first integral in~(\ref{eqn:fullflow_M}) (the other purely fermionic loop integrals are performed analogously). We have  

\begin{align}
\partial_\Lambda^R\int_k G(k)^2 \sin^2\theta&=\int_k G(k) S_f(k)\nonumber\\&=
k_F\int_{-\kUV}^{\kUV}\frac{dl}{2\pi}\int_{-\infty}^{\infty}\frac{d\omega}{2\pi}\frac{\omega^2}{\omega^2+\Lambda^2}\frac{1}{i\Aomega\omega-\Ak l}\left[-\frac{\omega^2\Lambda}{(\omega^2+\Lambda^2)^2}\frac{i(2-\eta_\omega)\Aomega\omega-(2-\eta_\mathbf{k})\Ak l}{(i\Aomega\omega-\Ak l)^2}\right]\nonumber\\&=-\frac{k_F}{\Ak\Aomega\Lambda}\int_{-Y}^Y\frac{db}{2\pi}\int_{-\infty}^{\infty}\frac{da}{2\pi}\frac{a^4}{(a^2+1)^3}\frac{i(2-\eta_\omega)a-(2-\eta_\mathbf{k})b}{(ia-b)^3}~,
\end{align}
where in the last line we have changed variables to $a=\omega/\Lambda$ and $b=\Ak l/ (\Aomega \Lambda)$. The integral can be performed exactly, and the asymptotic form at large $Y$ (which obtains in this paper) is
\begin{align}
\frac{3k_F(2-\eta_\mathbf{k})}{16\pi\Ak\Aomega\Lambda Y}~.
\end{align}
Substituting into~(\ref{eqn:fullflow_M}) and rewriting in terms of $N=k_F/\kUV$ and the rescaled couplings~(\ref{eqn:rescaled_couplings}) leads to the first term in the beta function for $\delta$, in Eqn.~(\ref{eqn:beta_functions}).

As an example of a purely bosonic integral, the second integral in~(\ref{eqn:fullflow_lambda}) is 
\begin{align}
\partial_\Lambda^R\int_q D_{ij}(q)D^{ij}(q)&=2\int_q \frac{\BAomega\Lambda(\etaAomega-2)}{\left(\BAomega(\Omega^2+\Lambda^2)+\BAq|\mathbf{q}|^2+M_\mathbf{A}^2\right)^3}\nonumber\\&=\frac{(\etaAomega-2)\Aomega^2}{16\pi\BAomega^2\Lambda^2\Ak^2\zeta^2}\left[\frac{1}{(1+\delta)^{3/2}}-\frac{1}{(Y^2\zeta^2+\delta+1)^{3/2}}\right]~.
\end{align}
\\\vspace{2mm}
\begin{center}
\textbf{2. Flow equations with mixed bosonic and fermionic loops}
\end{center}

The flow equations for $\Aomega,\Ak$ and $\tilde{g}_\mathbf{A}$ feature diagrams that have both internal fermionic and bosonic lines, and are somewhat more difficult to evaluate. We shall assume $N=k_F/\kUV \gg 1$, since (a) this is the most physical limit (as mentioned in the main text) and (b) the loop integrals are not tractable otherwise. The flow equation for $\Aomega$ is 
\begin{equation}
\partial_\Lambda \Aomega = i\frac{\partial}{\partial \omega}\bigg\rvert_{\omega=0,|\mathbf{k}|=k_F}\left[\partial_\Lambda^R\int_{k'}G(k')\Gamma^{(2,1)}_i(k,k',k-k')\Gamma^{(2,1)}_j(k',k,k'-k)D^{ij}(k-k')\right]~.
\end{equation}
The flow equation for $\Ak$ is
\begin{equation}
\label{eqn:fullflow_Ak}
\partial_\Lambda \Ak = \frac{\partial}{\partial l}\bigg\rvert_{\omega=0,|\mathbf{k}|=k_F}\left[\partial_\Lambda^R\int_{k'}G(k')\Gamma^{(2,1)}_i(k,k',k-k')\Gamma^{(2,1)}_j(k',k,k'-k)D^{ij}(k-k')\right]~.
\end{equation}
The flow equation for $\tilde{g}_\mathbf{A}$ is
\begin{align}
\label{eqn:fullflow_g}
\partial_\Lambda \left(\tilde{g}_\mathbf{A}\hat{k}_i\right)= -\partial_\Lambda^R\int_{k'}G(k')^2 D^{jl}(k'-k) \Gamma^{(2,1)}_i(k',k',0)\Gamma^{(2,1)}_j(k',k,k'-k)\Gamma^{(2,1)}_l(k,k',k-k')~,
\end{align}
where we set $\omega=0$ and $|\mathbf{k}|=k_F$.

Now, as argued in the main text the Yukawa coupling $\Gamma^{(2,1)}(k,k',k-k')$ is irrelevant except at small momentum transfers. By ``small'', we mean $|\mathbf{k}-\mathbf{k'}|\ll k_F$. Let $\Delta k'_\perp$ and $\Delta k'_\parallel$ be the components of $\mathbf{k}'-\mathbf{k}$ that are perpendicular and parallel to $\mathbf{k}$, respectively. For $\mathbf{k'}$ close to $\mathbf{k}$, $\Delta k'_\parallel$ is at most of the order of $\kUV$, which is $\ll k_F$ as $N\gg 1$. So, to satisfy the smallness condition we need $|\Delta k'_\perp|\ll k_F$, which is achieved by taking $|\Delta k'_\perp|\lesssim \kUV$. For $\mathbf{k'}$ close to $\mathbf{k}$ we have $\Delta k'_\perp\approx k_F \theta'$, where $\theta'$ is the angular variable for $\mathbf{k}'$ in the above integrals, measured from the direction of $\mathbf{k}$. We thus have $|\theta'|\lesssim \kUV/k_F$, i.e.\ we must restrict the limits of the angular integration to $\pm 1/N$. The variables appearing in the integration are depicted in Fig.~\ref{fig:ang_region}.

\begin{figure}
\centering
\includegraphics[width=0.45\linewidth]{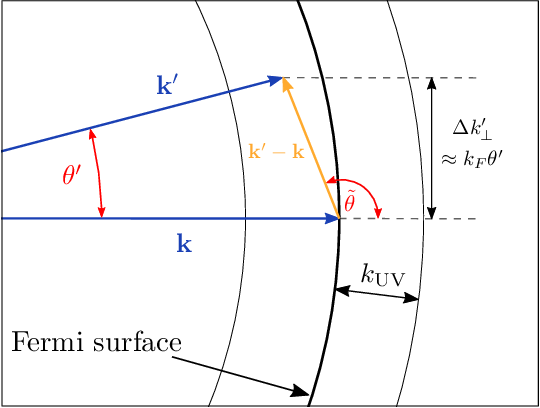}
\caption{The kinematic variables appearing in loop integrals containing internal bosonic and fermionic lines. $\mathbf{k}$ lies on the Fermi surface and is the external momentum; $\mathbf{k}'$ is integrated over a region around $\mathbf{k}$ of area $\mathcal{O}(\kUV^2)$.}
\label{fig:ang_region}
\end{figure}

Let us now evaluate~(\ref{eqn:fullflow_g}) --- the other flow equations are analogous. Since $|\mathbf{k}-\mathbf{k'}|\lesssim \kUV\ll k_F$, we shall approximate $\hat{\mathbf{k}}'\approx \hat{\mathbf{k}}$ in the Yukawa couplings (we cannot do this in the propagators, since they tend to non-smooth functions of $\mathbf{k'}$ as $\Lambda\to 0$) --- that is,
\begin{align}
\partial_\Lambda \tilde{g}_\mathbf{A}&\approx \tilde{g}_\mathbf{A}^3\partial_\Lambda^R\int_{k'}G(k')^2D^{jl}(k'-k)\hat{k}_j\hat{k}_l\nonumber\\&=\tilde{g}_\mathbf{A}^3\int_{k'}\left[2~G(k')S_f(k')D^{jl}(k'-k)+G(k')^2S_\mathbf{A}^{jl}(k'-k)\right]\hat{k}_j\hat{k}_l~.
\end{align}
The index structure in either term on the last line gives rise to a factor
\begin{align}
\left(g^{jl}+\frac{(k'-k)^j(k'-k)^l}{(\mathbf{k}'-\mathbf{k})^2}\right)\hat{k}_j \hat{k}_l=-1+\frac{\left[\hat{\mathbf{k}}\cdot(\mathbf{k}'-\mathbf{k})\right]^2}{(\mathbf{k}'-\mathbf{k})^2}=-\sin^2\tilde{\theta}~,
\end{align}
where $\tilde{\theta}$ is the angle between $\mathbf{k}'-\mathbf{k}$ and $\mathbf{k}$ (distinct from $\theta'$ --- see Fig.~\ref{fig:ang_region}). Additionally, in the denominators of the boson propagators $D$ and $S_\mathbf{A}$ there will be a term proportional to $(\mathbf{k}'-\mathbf{k})^2$. Since $\mathbf{k}$ is on the Fermi surface, we have
\begin{align}
(\mathbf{k}'-\mathbf{k})^2&=2k_F^2(1-\cos\theta')+2k_F l' (1-\cos\theta')+l'^2\nonumber\\&\approx k_F^2 \theta'^2+k_Fl'\theta'^2+l'^2~,
\end{align}
where we used $|\theta'|\ll 1$ in the last line. The first and last terms above are $\mathcal{O}(\kUV^2)$, while the middle term is $\mathcal{O}(\kUV^2/N)$, so we neglect it. Putting all this together, we get
\begin{align}
\partial_\Lambda \tilde{g}_\mathbf{A}\approx \ &\tilde{g}_\mathbf{A}^3k_F\int_{-1/N}^{1/N}\frac{d\theta'}{2\pi}\int_{-\kUV}^{\kUV}\frac{dl'}{2\pi}\int_{-\infty}^{\infty}\frac{d\omega'}{2\pi}\frac{\omega'^2}{\omega'^2+\Lambda^2}\frac{1}{i\Aomega\omega'-\Ak l'}\times\nonumber\\&\times\left[-2\frac{\omega'^2\Lambda}{(\omega'^2+\Lambda^2)^2}\frac{i(2-\eta_\omega)\Aomega\omega'-(2-\eta_\mathbf{k})\Ak l'}{(i\Aomega\omega'-\Ak l')^2}\frac{\sin^2\tilde{\theta}}{\BAomega(\omega'^2+\Lambda^2)+\BAq(k_F^2\theta'^2+l'^2)+M_\mathbf{A}^2}\right.\nonumber\\&\left.-\frac{\omega'^2}{\omega'^2+\Lambda^2}\frac{1}{i\Aomega\omega'-\Ak l'}\frac{\BAomega\Lambda(2-\etaAomega)\sin^2\tilde{\theta}}{\left(\BAomega(\omega'^2+\Lambda^2)+\BAq(k_F^2\theta'^2+l'^2)+M_\mathbf{A}^2\right)^2}\right]~.
\end{align}

Then, we introduce the variables $b=l'\Ak/(\Aomega \Lambda)$ and $c=k_F\theta'\Ak/(\Aomega \Lambda)$, dimensionless Cartesian coordinates about $\mathbf{k}$. We also use $a=\omega'/\Lambda$. $b$ ($c$) is the coordinate parallel (perpendicular) to $\hat{\mathbf{k}}$, and they both range from $-Y$ to $Y$ in the integral. Next, we transform to circular polar coordinates $b=\rho\cos\tilde{\theta}$ and $c=\rho\sin\tilde{\theta}$, and approximate the square integration region about $\mathbf{k}$ by a circle, to get
\begin{align}
\partial_\Lambda \tilde{g}_\mathbf{A}\approx \frac{\tilde{g}_\mathbf{A}^3}{\Ak^2 \Lambda^2 \BAomega}\int_{-\infty}^{\infty} \frac{da}{2\pi}\int_{0}^Y\frac{d\rho~\rho}{2\pi}\int_0^{2\pi}\frac{d\tilde{\theta}}{2\pi}\left[-2\frac{a^4}{(a^2+1)^3}\frac{i(2-\eta_\omega)a -(2-\eta_\mathbf{k})\rho\cos\tilde{\theta}}{\left(ia-\rho\cos\tilde{\theta}\right)^3}\frac{\sin^2\tilde{\theta}}{a^2+1+\zeta^2\rho^2+\delta}\right.\nonumber\\\left.+(\etaAomega-2)\left(\frac{a^2}{a^2+1}\right)^2\frac{1}{\left(ia-\rho\cos\tilde{\theta}\right)^2}\frac{\sin^2\tilde{\theta}}{(a^2+1+\zeta^2\rho^2+\delta)^2}\right]~.
\end{align}
The integrals over $\tilde{\theta}$ can be done, leaving double integrals over $a$ and $\rho$, which correspond to the functions $G_i$ in Appendix~\ref{append:integrals}:
\begin{align}
\partial_\Lambda \tilde{g}_\mathbf{A} \approx \frac{\tilde{g}_\mathbf{A}^3}{\Ak^2 \Lambda^2 \BAomega}\biggr[-2i(2-\eta_\omega)G_2(\delta,\zeta,Y)+2(2-\eta_\mathbf{k})G_3(\delta,\zeta,Y)+(\etaAomega-2)G_1(\delta,\zeta,Y)\biggr]~.
\end{align}
Rewriting in terms of the rescaled couplings produces $\eta_{\tilde{g}}$ in~(\ref{eqn:anom}).
\\\vspace{2mm}
\begin{center}
\textbf{3. Modified Ward-Takahashi identities}
\end{center}

As for the flow equations, we substitute the ansatz~(\ref{eqn:Ansatz}) into the master mWTI~(\ref{eqn:mastermWTI}) and compare coefficients of monomials in the fields. The right-hand side of the master identity is treated similarly to that of the Wetterich equation. First, the term $\frac{\delta^2\mathcal{G}}{\delta\eta_\sigma(k)\delta\etabar_\sigma(k+q)}\big\rvert_{\eta^*,\etabar^*,\mathbf{J}^*}$ on the right-hand side is an ``entry'' of $\mathbf{G}^{(2)}[\eta^*,\etabar^*,\mathbf{J}^*]$, the infinite-dimensional matrix of second derivatives of $\mathcal{G}$ with respect to the source fields, where $\eta^*,\etabar^*$ and $\mathbf{J}^*$ are those source fields satisfying $\psibar=\delta\mathcal{G}/\delta\eta\rvert_{\eta^*,\etabar^*,\mathbf{J}^*}$, etc. We then have the reciprocity relation for second derivatives of Legendre transforms~\cite{WMMSCH2012}, 
\begin{align}
\mathbf{G}^{(2)}[\eta^*,\etabar^*,\mathbf{J}^*]=\left(\mathbf{\Gamma}^{(2)}[\psi,\psibar,\mathbf{A}]+\mathbf{R}\right)^{-1}=\left(\mathbf{G}^{-1}-\tilde{\mathbf{\Sigma}}[\psi,\psibar,\mathbf{A}]\right)^{-1}~.
\end{align}
Like in the Wetterich equation, this inverse can then be expanded in a Neumann series and coefficients of monomials in fields can be matched. The identities considered in this paper are
\begin{eqnarray}
q_\mu\Pi^{\mu\nu}(q) & = & -2e\int_k\left[R_f(k+q)-R_f(k)\right]G(k)G(k+q)\Gamma^{(2,1)\nu}(k+q,k,q)~,\nonumber\\
q_{\mu_1}\Gamma^{(0,4)\mu_1\mu_2\mu_3\mu_4}(-q,q_2,q_3,q-q_2-q_3) & = & 12e\int_k\left[R_f(k+q)-R_f(k)\right]\mathcal{S}_{234}\biggr(G(k)G(k+q_2)G(k+q_2+q_3)\times \nonumber \\ & & \qquad \times \,G(k+q_2+q_3+q_4)\Gamma^{(2,1)\mu_2}(k+q_2,k,q_2) \times \phantom{\big\rvert} \nonumber \\
& & \qquad \times\,\Gamma^{(2,1)\mu_3}(k+q_2+q_3,k+q_2,q_3) \times \phantom{\big\rvert}\nonumber \\
& & \qquad \times\, \Gamma^{(2,1)\mu_4}(k+q_2+q_3+q_4,k+q_2+q_3,q_4)\biggr)\bigg\rvert_{q_4=q-q_2-q_3}~,
\end{eqnarray}
along with (\ref{eqn:Yukawa_mWTI}). In the above, $\mathcal{S}_{234}$ denotes symmetrization in $(q_2,\mu_2)$, $(q_3,\mu_3)$ and $(q_4,\mu_4)$. 

The loop integrals are treated analogously to those in the flow equations. To demonstrate that the mWTIs imply $\Gamma^{(2,1)i}(\hat{\mathbf{k}})\propto\hat{k}^i$, as claimed below (\ref{eqn:Yukawa_form}), let us briefly consider~(\ref{eqn:Yukawa_mWTI}). Taking the derivative $\partial/\partial q_i\rvert_{q=0}$ and setting $\omega=0$, $|\mathbf{k}|=k_F$, we get
\begin{align}
\Gamma^{(2,1)i}(k,k,0)=-e\Ak\hat{k}^i+e\int_{k'}\Ak (\hat{k}')^i\frac{\Lambda^2}{\omega'^2} G(k')^2D^{\mu\nu}(k'-k)\Gamma^{(2,1)}_\mu(k',k,k'-k)\Gamma^{(2,1)}_\nu(k,k',k-k')~.
\end{align}
One might also expect terms on the left-hand side and in the integral that correspond to acting with $\partial/\partial q_i$ on propagators or vertex functions that have $q$ in the argument. Such terms give zero, as the derivatives are regular at $q=0$ and multiply a factor that vanishes at $q=0$. Then, as for the flow equations we restrict the angular integration limits to $\pm 1/N$, and consequently we can approximate the factor $(\hat{k}')^i$ inside the integral by $\hat{k}^i$ (the error induced by this is $\mathcal{O}(1/N)$). Thus, the Yukawa coupling is proportional to $\hat{\mathbf{k}}$.

\end{widetext}

\end{appendices}

%




\end{document}